\documentclass[a4paper,11pt]{article}
\usepackage{jheppub} % for details on the use of the package, please see the JINST-author-manual
\usepackage{lineno}
\makeatletter
\gdef\@fpheader{}
\makeatother

\def\sqr#1#2{{\vcenter{\vbox{\hrule height.#2pt
            \hbox{\vrule width.#2pt height#1pt \kern#1pt
                  \vrule width.#2pt}\hrule height.#2pt}}}}

\def\square
 {\mathop{\mathchoice{\sqr{12}{15}}{\sqr{9}{12}}
{\sqr{6.3}{9}}{\sqr{4.5}{9}}}}

\def\sqra#1#2#3{{\vcenter{\vbox{\hrule height.#2pt
            \hbox{\vrule width.#2pt height#1pt \kern5pt %\kern#1pt
#3
                  \vrule width.#2pt}\hrule height.#2pt}}}}

\title{Decomposition squared}

\author{E.~Sharpe,}
\affiliation{Physics Department, Virginia Tech\\
850 West Campus Drive, Blacksburg, VA. 24061}
\emailAdd{ersharpe@vt.edu}

\author{H.~Zhang}
\emailAdd{hzhang96@vt.edu}

\abstract{In this paper, we test and extend a proposal of Gu, Pei, and Zhang for an application of decomposition to three-dimensional theories with one-form symmetries and to quantum K theory.  The theories themselves do not decompose, but, OPEs of parallel one-dimensional objects (such as Wilson lines) and dimensional reductions to two dimensions do decompose, sometimes in two independent ways.  We apply this to extend conjectures for quantum K theory rings of gerbes (realized by three-dimensional gauge theories with one-form symmetries) via both orbifold partition functions and gauged linear sigma models.  
}

\begin{document}
\maketitle
\flushbottom

\section{Introduction}

Decomposition is the observation that a local $d$-dimensional quantum field theory with a global $(d-1)$-form symmetry is equivalent to (``decomposes'' into) a disjoint union of local theories without global $(d-1)$-form symmetries.
Decomposition was first described in 2006 in \cite{Hellerman:2006zs} as part of efforts to resolve some of the technical difficulties in making sense of string propagation on stacks and gerbes \cite{Pantev:2005rh,Pantev:2005wj,Pantev:2005zs}, and it has been documented and confirmed since in many examples, many kinds of examples, in different dimensions,
see for example \cite{Caldararu:2007tc,Anderson:2013sia,Sharpe:2014tca,Sharpe:2019ddn,Tanizaki:2019rbk,Komargodski:2020mxz,Gu:2020ivl,Eager:2020rra,Cherman:2020cvw,Nguyen:2021yld,Nguyen:2021naa,Gu:2021yek,Gu:2021beo,Robbins:2021lry,Robbins:2021ibx,Robbins:2021xce,Sharpe:2021srf,Pantev:2022kpl,Pantev:2022pbf,Lin:2022xod,Robbins:2022wlr,Perez-Lona:2023llv,Pantev:2023dim,Perez-Lona:2023djo,Sharpe:2023lfk,Bhardwaj:2024qrf,Honda:2021ovk,Meynet:2022bsg} for a sample of the literature, and see also \cite{Sharpe:2010zz,Sharpe:2010iv,Sharpe:2019yag,Sharpe:2022ene} for introductions and reviews. 

One of the original predictions of decomposition \cite{Hellerman:2006zs} was that the Gromov-Witten invariants and
quantum cohomology rings of
gerbes should be equivalent to the those of disjoint unions of spaces.
Gerbes are essentially bundles of one-form symmetries, hence a two-dimensional
sigma model whose target space is a gerbe should admit 
a global one-form symmetry (corresponding to translation along the fibers of the target), and so decompose.  This prediction for Gromov-Witten theory was checked in e.g.~\cite{ajt1,ajt2,ajt3,t1,gt1,xt1}.

In this paper, we extend a proposal of \cite[section 4]{Gu:2021yek}, \cite[section 3]{Gu:2021beo}, and discuss extensions of such notions to three-dimensional theories with one-form symmetries, and corresponding predictions of decomposition for quantum K theory.  In particular,
quantum K theory rings can be computed using three-dimensional gauge theories  
\cite[section 2.4]{Bullimore:2014awa}, \cite{Jockers:2018sfl,Jockers:2019wjh,Jockers:2019lwe,Ueda:2019qhg,Nekrasov:2009uh,Closset:2016arn,Closset:2017zgf,Closset:2018ghr,Closset:2019hyt,Closset:2023vos,Closset:2023bdr,Gu:2020zpg,Gu:2022yvj,Gu:2023tcv,Gu:2023fpw},
and for the same reasons as above,
the three-dimensional gauge theories for gerbes
have one-form symmetries.  Now, such three-dimensional theories themselves do not decompose, as that would
require a two-form symmetry.  (A three-dimensional sigma model whose target is a 2-gerbe, on the other hand,
would have a global two-form symmetry and so decompose.)
However, the quantum K theory rings are computed as OPEs of unlinked parallel Wilson lines, wrapped on the same $S^1$, which leads to two parallel effects, potentially two different decompositions:
\begin{itemize}
    \item Those Wilson lines can be acted upon by the generator of the global one-form symmetry to produce other Wilson lines, so that there is a multiplicity, which results in one decomposition.
    \item For much the same reasons that electromagnetism of infinite parallel planes in three dimensions reduces to an effectively one-dimensional problem, here the pertinent aspects of the three-dimensional theory are captured by an effective two-dimensional theory.  Each such two-dimensional theory has a one-form symmetry, and so decomposes.
\end{itemize}
More technically, given a $BK$ (one-form) symmetry in three dimensions, after Kaluza-Klein reduction on $S^1$, the two-dimensional theory has a $BK \times K$ symmetry:
\begin{itemize}
    \item Wilson lines wrapped along the $S^1$ generate dimension-zero defects (and the one-form symmetry $BK$) in the two-dimensional theory,  This leads to one level of decomposition in two dimensions.
    \item Real-codimension-one defects perpendicular to $S^1$ (corresponding, as we will argue later, to the zero-form symmetry $K$) couple to states in copies of the two-dimensional space.  (In orbifold constructions, these are twisted sectors along the $S^1$.)  These are invisible in a dimensional reduction, but appear in a more complete Kaluza-Klein reduction.
    These are better understood as superselection sectors rather than decomposition, but for the purpose of understanding IR phenomena such as quantum K theory, the effect is very similar.
\end{itemize}
As a result, deep in the IR, there are potentially two different notions of decomposition operating in such examples.  If there is no 't Hooft anomaly between the two-dimensional $BK$ and $K$ (in three dimensions, if there is no self-'t Hooft anomaly in the $BK$), then both mechanisms operate independently, so that for a (banded) ${\mathbb Z}_{\ell}$ gerbe, deep in the IR, one gets $\ell^2$ universes, as first remarked in \cite[section 4]{Gu:2021yek}, a result we will see explicitly in both orbifold partition functions and also in physics computations of quantum K theory rings.

In particular, we will use three-dimensional gauge theories to justify and illustrate the conjecture that the quantum K theory ring of a gerbe is equivalent to the quantum K theory ring of a disjoint union (of squared order), as expected from the physics of decomposition. 

We should emphasize that this phenomenon is not specific to one-form symmetries in
three-dimensional theories.  For example, schematically,
given a $d$-dimensional theory with a ${\mathbb Z}_{\ell}$ $(d-k-1)$-form
symmetry, say, one can construct projection operators on parallel $k$-dimensional objects, and by doing a Kaluza-Klein reduction along a $k$-dimensional factor in the spacetime manifold, one potentially gets a disjoint union of $\ell$ low-energy theories of dimension $d-k$ with a ${\mathbb Z}_{\ell}$ $(d-k-1)$-form symmetry, each of which separately decomposes, for a total of $\ell^2$ universes.  (To get this additional structure assumes no mixed 't Hooft anomalies, and also may require, on dimensional grounds, that $d-k-1 \leq k$, or $2k \geq d-1$, so that the $(d-k-1)$-form symmetry may reduce to a zero-form symmetry.)  Other variations also exist, and will be discussed in future work.  Related ideas have also appeared in discussions of compactifications of six-dimensional (2,0) theories, see for example
\cite[section 2.1]{Chun:2019mal}.

We should also emphasize that to see this phenomenon requires keeping track of
modes wrapped on the $S^1$.  In other words, the point of this paper is to
discuss a phenomenon arising in Kaluza-Klein reductions.  By contrast, in
a dimensional reduction on an $S^1$, when all dependence on the $S^1$ is
merely truncated, we do not expect these phenomena to arise.

We begin in section~\ref{sect:predict} by making a prediction for dimensional reductions and OPEs of parallel one-dimensional objects in three-dimensional $G$ gauge theories with trivially-acting $K \subset G$ (and hence a one-form symmetry).  The previous papers \cite{Gu:2021yek,Gu:2021beo} considered the case that $K$ is a subset of the center of $G$, and we extend the proposal to more general $K$, not necessarily central -- meaning, not-necessarily-banded gerbes.  In such more general cases, the statement of decomposition is more complex than in cases in which $K$ is central.

In section~\ref{sect:orb} we discuss that prediction in the case of global orbifolds by finite groups.  Our construction of the dimensional reduction explicitly reproduces the form of the prediction of section~\ref{sect:predict}, but we think it useful to illustrate the consequences in a number of different kinds of examples.

In section~\ref{sect:glsm} we turn to gauged linear sigma model (GLSM) computations.  In the global orbifolds of the previous section, we could only discuss the form of the decomposition (the disjoint union), but not the quantum K theory rings of the separate universes.  Using GLSM methods, we are able to discern both the decomposition as well as the quantum K theory rings of the individual universes.  In each case we discuss the quantum cohomology of a two-dimensional GLSM and the quantum K-theory ring from a three-dimensional GLSM, and in each case, discuss how the decomposition can be seen explicitly.  Specifically, in subsection~\ref{sect:proj}, we review gerbes on projective spaces, for which results already exist in the literature.
In subsection~\ref{sect:ggerbe:gkn} we discuss general ${\mathbb Z}_{\ell}$ gerbes on Grassmannians $G(k,n)$, and in section~\ref{sect:wgerbe}, we turn to ${\mathbb Z}_{\ell}$ gerbes presented as weighted Grassmannians, analogues of weighted projective spaces.  Mathematically, these are special cases of the gerbes in the previous section, but their physical presentation is different, so we repeat the analysis here, and outline how the physical predictions for quantum K theory rings are consistent with expectations of decomposition.

For completeness, in section~\ref{sect:genl-wgrass} we discuss general weighted Grassmannians and predictions for their quantum K theory rings.
In section~\ref{sect:gerbe:flag} we perform the same analyses for ${\mathbb Z}_{\ell}$ gerbes on flag manifolds.

The same methods can be applied to study other spaces beyond those above -- for example, gerbes on Fano toric varieties.  However, the methods and analyses are essentially the same as that discussed here, so for the purposes of this paper, we feel that the examples above should suffice to set up the conjecture that quantum K theory of gerbes is equivalent to quantum K theory of disjoint unions of spaces, via longitudinal decomposition.

In appendix~\ref{app:bundles-gerbes} we discuss bundles on stacks and gerbes,
as relevant for discussions of quantum K theory.

In passing, the fact that dimensional reduction can yield disjoint unions plays an essential role in this paper, and has also been discussed in a different context in \cite{Aharony:2017adm}.

\section{Prediction} \label{sect:predict}

First, we briefly recall decomposition in two-dimensional gauge theories, before turning to
three-dimensional examples.  Consider a two-dimensional $G$ gauge theory (which we denote
$[X/G]$, in obvious reference to orbifolds, but the idea holds more generally) in which
a subgroup $K$ acts trivially.  This theory has\footnote{
Strictly speaking, one speaks of higher-form symmetries only for abelian groups.  However,
decomposition is slightly more general -- there is a decomposition even if the trivially-acting
group is nonabelian.
} a one-form symmetry, and so one expects a decomposition.
Then \cite{Hellerman:2006zs} 
\begin{equation}
    {\rm QFT}_{\rm 2d}\left( [X/G] \right) \: = \: {\rm QFT}_{\rm 2d} \left( \left[ \frac{ X \times \hat{K} }{G/K} \right]_{\omega} \right),
\end{equation}
where $\hat{K}$ denotes the set of irreducible representations of $K$, and $\omega$ denotes
discrete torsion described in \cite{Hellerman:2006zs}.  In the special case that the effectively-acting
group $G/K$ acts trivially on $\hat{K}$, the right-hand-side becomes a disjoint union of $G/K$ gauge
theories, as many as irreducible representations of $K$.  

Next, consider a three-dimensional $G$ gauge theory, again denoted $[X/G]$, with a trivially-acting
subgroup $K \subset G$.  For the moment, we assume the theory does not have a Chern-Simons term, and return to such terms later.
This theory has a one-form symmetry, but in three dimensions, this does not predict a decomposition.  However, if we consider either a dimensional reduction to two dimensions,
or alternatively consider a theory of parallel one-dimensional objects (such as Wilson lines, as relevant to quantum K theory), then the low-energy effective two-dimensional theory decomposes,
and we predict that at low energies, decomposition has the form
\begin{equation}
    {\rm QFT}_{\rm 3d \, eff'} \left( [X/G] \right) \: = \: {\rm QFT}_{\rm 2d} \left( \coprod_{[g]} \left[ \frac{ X \times \hat{K}_g }{ C(g)/K_g } \right]_{\omega} \right),
\end{equation}
where the disjoint union is over trivially-acting conjugacy classes $[g]$ of $G$, $C(g) \subset G$ denotes the
centralizer of a representative $g \in G$, $K_g \subset C(g)$ is the trivially-acting subgroup of the centralizer, and finally $\omega$ denotes discrete torsion, the same discrete torsion that would arise
in a two-dimensional $C(g)$ gauge theory with the same matter, as described in
\cite{Hellerman:2006zs}.

As an important special case, suppose that the trivially-acting subgroup of the (original) group $G$ lies within the center of $G$: $K \subset Z(G) \subset G$.  Then, the set of trivially-acting conjugacy classes is equivalent to the set of elements of $K$, and the low-energy decomposition above reduces
to the statement that 
\begin{equation}
    {\rm QFT}_{\rm 3d \, eff'}\left( [X/G] \right) \: = \: {\rm QFT}_{\rm 2d}\left( \coprod_{g \in K} \coprod_{\rho \in \hat{K}} [X / (G/K) ]_{\omega(\rho)} \right),
\end{equation}
which has $|K|^2$ universes\footnote{
As is discussed elsewhere, this is due to a combination of decomposition and superselection sectors, so deep in the IR this is $|K|^2$ universes, but should be more invariantly understood
as a combination of decomposition and superselection sectors rather than just decomposition per se.
}, rather than $|K|$ as in the analogous two-dimensional case.
This special case was discussed in \cite{Gu:2021yek,Gu:2021beo}; part of the point of this paper is to extend that to more general cases.

We can understand this as a consequence of two orthogonal effects, both arising from the one-form symmetry $BK$ of the gauge theory in three dimensions, on a three-manifold of the form $S^1 \times \Sigma$:
\begin{itemize}
    \item The line operators for the $BK$ along the $S^1$ reduce to pointlike operators on $\Sigma$, and so reduce to a one-form symmetry on $\Sigma$, responsible for one decomposition.
    \item In addition, there are twisted sector states along the $S^1$, arising in the gauge theory.  In three dimensions, twisted sector states are supported along two-dimensional surfaces, whose intersection with $\Sigma$ corresponds to line operators for $K$ in the
    two-dimensional theory, or equivalently line operators for $BK$ in the three-dimensional theory.  Strictly speaking, since these do not arise from a separate one-form symmetry but rather a zero-form symmetry, it is better to understand the resulting sectors as superselection sectors.
\end{itemize}
As a result, schematically the theory has a 
\begin{equation}
    \left( \mbox{decomposition} \right) \times \left( \mbox{superselection sectors} \right)
\end{equation}
structure, rather than a
\begin{equation}
    \left( \mbox{decomposition} \right)^2
\end{equation}
structure per se.  However, much of our interest will focus on quantum K theory and other IR effects for which the distinction is moot (hence our focus on low-energy behaviors). 

So far, we have discussed theories without Chern-Simons terms.  A Chern-Simons term will modify the one-form symmetry, and hence the structure of the decomposition.  
We compare several cases to illustrate this.  We focus on abelian theories, both as prototypes for more general cases, and also because in GLSM computations, generically on the Coulomb branch the gauge symmetry is abelian.
\begin{itemize}
\item First, consider a three-dimensional $U(1)$ Chern-Simons theory at level $m$,
a $U(1)_m$ theory.  This theory has a $B {\mathbb Z}_m$ (one-form) symmetry, with line operators given by Wilson lines of various charges.  
If $m$ is even, there are $m$ distinct line operators of the form
\begin{equation}
    W_n \: = \: \exp\left( i n \oint A \right),
\end{equation}
related by
$n \cong n \mod m$, often conventionally labelled \cite[appendix C.1]{Seiberg:2016rsg}
\begin{equation}
    n \: = \: 0, \pm 1, \cdots, \pm \frac{m-2}{2}, + \frac{m}{2}.
\end{equation}
If $m$ is odd, the theory can be defined only if the underlying three-manifold is spin,
and there are $2m$ line operators of the form above ($n$ is no longer quite equivalent to
$n+m$).
(See \cite[appendix C]{Seiberg:2016rsg}, \cite[appendix A]{Seiberg:2016gmd}, \cite[appendix C]{Hsin:2018vcg}, \cite[section 2.2]{Benini:2022hzx}, \cite[section 5.9]{Pantev:2022pbf}, \cite{Belov:2005ze,Freed:1992vw,freed2} for more information.)

For more information on the identifications above, see for example \cite{stackexchange}.

\item A three-dimensional $U(1)$ gauge theory with matter fields of charges all multiples of $k$,
and no Chern-Simons term, has a $B {\mathbb Z}_k$ (one-form) symmetry.  In this case, the periodicity arises because a Wilson line $W_k$ can end on a field of charge $k$, so we can use those perturbative fields to `break' Wilson lines, so that $W_k \cong 1$.  Since we can write any $W_n = W_{n-k} \otimes W_k$, this results in a periodicty 
$W_n \cong W_{n - k}$.
\item Next, we combine these cases.  Consider a $U(1)$ gauge theory with matter fields of charges all multiples of $k$, and also with a Chern-Simons term at level $m$.  This theory has a $B {\mathbb Z}_{ {\rm gcd}(m,k) }$
(one-form) symmetry.  
To see this, we use B\'ezout's identity, which says that there exist integers $a$, $b$ such that
\begin{equation}
    a m + b k \: = \: {\rm gcd}(m,k),
\end{equation}
and moreover, integral linear combinations of $m$ and $k$ are multiples of gcd$(m,k)$.
As a result, by using combinations of the two periodicity mechanisms above, the Wilson lines $W_n$ are only distinct for $n \mod {\rm gcd}(m,k)$.

For simplicity we restrict to the case $m$ is even.
Suppose $m=2$, $k=3$, which have gcd$(m,k) = 1$.  The distinct Wilson lines of the Chern-Simons theory at level $2$ have $n=0, 1$.  Now,
\begin{eqnarray}
n = 1 & \cong & n = 4 \mbox{ using the $k$ periodicity}, \\
    & \cong & n = 0 \mbox{ using the $m$ periodicity}, 
\end{eqnarray}
and so there are no nontrivial Wilson lines -- all are equivalent to the identity, as expected from the gcd.

For another example, suppose $m=6$, $k=4$, which have gcd$(m,k) = 2$.  The allowed Wilson lines of the Chern-Simons theory at level $6$ have 
\begin{equation}
    n \: = \: 0, \pm 1, \pm 2, + 3.
\end{equation}
Now,
\begin{eqnarray}
    n = 3 & \cong & n = 7 \mbox{ using the $k$ periodicity}, \\
    & \cong & n = 1 \mbox{ using the $m$ periodicity,} \\
    n = 2 & \cong & n = 6 \mbox{ using the $k$ periodicity,} \\
    & \cong & n = 0 \mbox{ using the $m$ periodicity,} \\
    n = -1 & \cong & n = 5 \mbox{ using the $m$ periodicity,} \\
    & \cong & n = 1 \mbox{ using the $k$ periodicity,} \\
    n = -2 & \cong & n = 4 \mbox{ using the $m$ periodicity,} \\
    & \cong & n = 0 \mbox{ using the $k$ periodicity.}
\end{eqnarray}
so that the $U(1)_6$ Chern-Simons theory effectively only has two distinct Wilson lines ($W_0$, $W_1$) in the
presence of charge 4 matter, as expected from the gcd.

\end{itemize}

As a result, in the presence of Chern-Simons terms, we must modify our prediction.
To further complicate matters, for $G$ gauge theories in which the trivially-acting subgroup $K \subset G$
is not abelian, the decomposition is not solely understandable in terms of one-form symmetries, as
$BK$ is only defined for $K$ abelian.  In this paper, in the presence of Chern-Simons terms, we only discuss
decomposition for $G$ gauge theories in which the trivially-acting subgroup is abelian.
(Decomposition will exist more generally, but we leave the matter of straightening out a precise prediction for future work.)

So far we have discussed conditions for the presence of a one-form symmetry that could generate one level of decomposition.  To get a second level of decomposition (or rather, independently operating superselection sectors), the two effects much act independently.  This means we must also require that the self-'t Hooft anomaly of that one-form symmetry in three dimensions, or equivalently the 't Hooft anomaly in two dimensions between the one-form symmetry and the corresponding reduced zero-form symmetry, vanish.

This 't Hooft anomaly was computed in, for example, \cite[section 5.1]{Closset:2024sle}.
For the level $m$ $U(1)$ Chern-Simons theory with matter of charge $k$ outlined above, it was argued that the 't Hooft anomaly is proportional to
\begin{equation}
    \frac{k}{m^2} \mod 1.
\end{equation}

Now, consider a $G$ gauge theory with trivially-acting abelian subgroup $K \subset Z(G) \subset G$.
Assume that, in the presence of Chern-Simons terms, there is a one-form symmetry $BL$ for $L \subset K$, and let us assume that there is no self-'t Hooft anomaly of $BL$ in three dimensions.
Then, in this case, we predict that 
\begin{equation}
    {\rm QFT}_{\rm 3d \, eff'}\left( [X/G] \right) \: = \: {\rm QFT}_{\rm 2d}\left( \coprod_{g \in L} \coprod_{\rho \in \hat{L}} [X / (G/L) ]_{\omega(\rho)} \right),
\end{equation}
which has $|L|^2$ universes.  Although a larger subgroup $K$ acts trivially, only $L \subset K$ will result in a decomposition, due to the presence of the Chern-Simons terms.
(We leave a systematic prediction for more general cases to future work.)

Now, we turn to quantum K theory, for a gerbe presented as a quotient $[X/G]$ where a subgroup $K \subset G$ acts trivially.  Quantum K theory is realized physically in a three-dimensional gauge theory on a three-manifold $\Sigma \times S^1$.
The quantum K theory ring is the OPE ring of parallel Wilson lines wrapped on the $S^1$, as discussed in \cite{Jockers:2018sfl,Jockers:2019wjh,Jockers:2019lwe,Closset:2017zgf,Closset:2018ghr,Closset:2019hyt,Closset:2023vos,Closset:2023bdr,Gu:2020zpg,Gu:2022yvj,Gu:2023tcv,Gu:2023fpw}.  In order to reproduce the quantum K theory ring appearing in mathematics, there are also Chern-Simons terms.  For a gauge theory with abelian gauge group and no superpotential, to match mathematics, the levels are given by \cite[section 2]{Gu:2023tcv}
\begin{equation}
    k^{ab} \: = \: - \frac{1}{2} (R_i - 1) \sum_i Q^i_a Q^i_b,
\end{equation}
where the $R_i$ denotes the $R$-charge of the $i$th chiral superfield, and
$Q^i_a$ is the charge of the $i$th chiral superfield under the $a$th $U(1)$ factor in a maximal torus of the gauge group.
(If all $R$ charges vanish, this reduces to the $U(1)_{-1/2}$ quantization described in 
e.g.~\cite[section 2.2]{Closset:2019hyt}.)

For our purposes, we observe that if there is a trivially-acting ${\mathbb Z}_k$ in the center of the gauge
group, then the charges $Q^i_a$ are divisible by $k$, and so the levels used in computing quantum K theory are divisible by $k$.  As a result, although quantum K theory is computed by a three-dimensional gauge theory with Chern-Simons terms, the Chern-Simons terms do not reduce any one-form symmetry arising from a subgroup of the gauge group acting trivially.  Furthermore, the levels are divisible by the square of the charges, so there is no 't Hooft anomaly.  In effect, for purposes of understanding decomposition, we can ignore the presence of the Chern-Simons terms.

Thus, for the quantum K theory of a gerbe presented as a quotient $[X/G]$ where a subgroup $K \subset G$ acts trivially, we predict
\begin{equation}
    {\rm QK}\left( [X/G] \right) \: = \: {\rm QK}\left( \coprod_{[g]} \left[ \frac{ X \times \hat{K}_g }{ C(g)/K_g } \right]_{\omega}  \right),
\end{equation}

Next, we shall check this prediction explicitly, in theories presented as global orbifolds
in section~\ref{sect:orb}, and in gauged linear sigma models in 
section~\ref{sect:glsm}.

\section{Presentations as global orbifolds}   \label{sect:orb}

Consider dimensional reductions of orbifolds $[X/G]$ from three dimensions to two-dimensions.  As this is a dimensional reduction, we omit dependence on the third dimension, hence we omit analogues of twisted sectors resulting from nontrivially-acting elements of $G$.  However, (conjugacy classes\footnote{Conjugacy classes, conjugating by elements of $G$, instead of group elements, because a gauge transformation will conjugate by elements of $G$.  As any trivially-acting subgroup is normal, conjugation will always map a conjugacy class to itself.} of ) trivially-acting elements of $G$ can still contribute along the third direction.  Then, for any one group element $g \in G$ (representing a conjugacy class) along the third direction, we are left with a two-dimensional orbifold by the centralizer $C(g)$.  As a result, we describe the partition function of a dimensionally-reduced orbifold $[X/G]$, on three-manifold $S^1 \times \Sigma$, as
\begin{equation}
    \sum_{[g]} [X/C(g)],
\end{equation}
where the sum is over conjugacy classes of $G$ that
act trivially on $X$.

Using known results for two-dimensional decompositions \cite{Hellerman:2006zs},
this immediately reproduces the structure of the three-dimensional decomposition
in section~\ref{sect:predict}.  In this section we will compute the result in a number of examples, to illustrate the range of phenomena that arise.  In each case, we will compute the partition function, after dimensional reduction, on a $T^2$.

In this section there will be no Chern-Simons terms in the three-dimensional theory to complicate
the analysis.  We shall consider examples with Chern-Simons terms later in
section~\ref{sect:glsm}.

In passing, in the special case that all of $G$ acts trivially, this has also been described in 
\cite{Qiu:2020opt,Muller:2020kos}, in discussions of dimensionally-reducing three-dimensional Dijkgraaf-Witten theory.  Analogous results in the condensed matter literature (in a different number of dimensions)
are also discussed in \cite[section 3.C]{Lan:2018vjb}.

\subsection{Example with central trivially-acting group}

Consider $[X/D_4]$, where $D_4$ is the eight-element dihedral group, with trivially-acting central ${\mathbb Z}_2 \subset D_4$, as in \cite[section 2.0.1]{Pantev:2005rh}, \cite[section 5.2]{Hellerman:2006zs}.  We can write
$D_4 = \langle a, b \rangle$ where
\begin{equation}
    a^2 = 1, \: \: \: b^2 = z, \: \: \: b^4 = 1,
    \: \: \: z^2 = 1, \: \: \:
    ba \: = \: abz,
\end{equation}
and $z$ generates the ${\mathbb Z}_2$ center.

Next, we will compute $T^2$ partition functions after dimensional reduction.

In either of the $1, z$ sectors (meaning, cases in which $1$, $z$ are inserted along the third $S^1$), the commuting pairs in $D_4$ which commute with that third group element form all of the commuting pairs in an ordinary two-dimensional orbifold, namely
\begin{equation}
    {\scriptstyle 1, z} \square_{1, z}, \: \: \:
    {\scriptstyle 1, z} \square_{a, az}, \: \: \:
    {\scriptstyle 1, z} \square_{b, bz}, \: \: \:
    {\scriptstyle 1, z} \square_{ab, ba}, \: \: \:
    {\scriptstyle a, az} \square_{1, z}, \: \: \:
    {\scriptstyle b, bz} \square_{1, z}, \: \: \:
    {\scriptstyle ab, ba} \square_{1, z}, \: \: \:
    {\scriptstyle a, az} \square_{a, az}, \: \: \:
    {\scriptstyle b, bz} \square_{b, bz}, \: \: \:
    {\scriptstyle ab, ba} \square_{ab, ba},
\end{equation}
which contribute
\begin{eqnarray}
\lefteqn{
    \frac{4}{|D_4|} \left[ Z_{1,1} + Z_{1,\overline{a}} + Z_{1, \overline{b}} + Z_{1, \overline{a} \overline{b}} + Z_{\overline{a}, 1} + Z_{\overline{b}, 1} + Z_{\overline{a} \overline{b}, 1} + Z_{\overline{a}, \overline{a}} + Z_{\overline{b}, \overline{b}} + Z_{\overline{a} \overline{b}, \overline{a} \overline{b}} \right]
    } \nonumber \\
    & \hspace*{1.5in} = & Z\left( [X/{\mathbb Z}_2 \times {\mathbb Z}_2] \, \coprod \, [ X/{\mathbb Z}_2 \times {\mathbb Z}_2]_{\rm d.t.} \right).
\end{eqnarray}
As there are two such sectors (one for each of $1$, $z$),
we see that the three-dimensional effective theory decomposes into
\begin{equation}
    \coprod_2 \left( [X/{\mathbb Z}_2 \times {\mathbb Z}_2] \, \coprod \, [ X/{\mathbb Z}_2 \times {\mathbb Z}_2]_{\rm d.t.} \right)
    \: = \:
    \coprod_2 [X/{\mathbb Z}_2 \times {\mathbb Z}_2] \, \coprod_2  [ X/{\mathbb Z}_2 \times {\mathbb Z}_2]_{\rm d.t.} .
\end{equation}
As observed earlier, this construction reproduces the
decomposition prediction of section~\ref{sect:predict}.

\subsection{First nonbanded example}

Consider the orbifold $[X/{\mathbb H}]$, where ${\mathbb H}$ is the eight-element group of unit quaternions $\{ \pm 1, \pm i, \pm j, \pm k\}$, with trivially-acting subgroup
$\langle i \rangle \cong {\mathbb Z}_4$, as in \cite[section 2.0.4]{Pantev:2005rh}, \cite[section 5.4.1]{Hellerman:2006zs}.  
The group ${\mathbb H}$ has conjugacy classes
\begin{equation}
    \{ 1 \}, \: \: \: \{ -1 \}, \: \: \:
    \{ \pm i \}, \: \: \:
    \{ \pm j \}, \: \: \:
    \{ \pm k \}.
\end{equation}

We construct the partition function on a torus by listing results for group elements along the
third $S^1$:
\begin{itemize}
    \item $\pm 1$: In each of these two sectors, all commuting pairs in ${\mathbb H}$ contribute:
    \begin{equation}
        {\scriptstyle \pm 1, \pm i} \square_{\pm 1, \pm i}, \: \: \:
        {\scriptstyle \pm 1} \square_{\pm j, \pm k}, \: \: \:
        {\scriptstyle \pm j, \pm k} \square_{\pm 1}, \: \: \: 
        {\scriptstyle \pm j} \square_{\pm j}, \: \: \:
        {\scriptstyle \pm k} \square_{\pm k}
    \end{equation}
    which contribute
    \begin{equation}
        \frac{1}{| {\mathbb H} | }\left[ (16) Z_{1,1} + (8) Z_{1,\xi} + (8) Z_{\xi,1} + (2)(4) Z_{\xi,xi}\right] 
        \: = \: Z\left( X \coprod_2 [X/{\mathbb Z}_2] \right),
    \end{equation}
    where $\xi$ represents the effectively-acting ${\mathbb Z}_2$, matching the usual result for decomposition in the corresponding two-dimensional orbifold.
    \item $\{ \pm i \}$:  In this sector, the allowed commuting pairs are
    \begin{equation}
         {\scriptstyle \pm 1, \pm i} \square_{\pm 1, \pm i},
    \end{equation}
    which contribute
    \begin{equation}
        \frac{1}{| C(i) |} \left[ (16) Z_{1,1} \right] 
        \: = \: (4) Z(X) 
        \: = \: Z\left( \coprod_4 X \right). 
    \end{equation}
\end{itemize}
There are no contributions from $\{ \pm j \}$, $\{ \pm k \}$ along the
third $S^1$, as they do not act trivially on $X$.  Altogether, the results from the $\{ 1 \}$, $\{ -1 \}$, $\{\pm i\}$ sectors are consistent with a decomposition into
\begin{equation}
    \coprod_6 X \, %\coprod \, 
    \coprod_4 [X/{\mathbb Z}_2].
\end{equation}
As observed earlier, this necessarily reproduces the prediction of
section~\ref{sect:predict}.

Note as a quick consistency check that if we think of each
$X$ as a double cover of $[X/{\mathbb Z}_2]$, then this is
locally 
\begin{equation}
    (6)(2) + 4 \: = \: 16 \: = \: 4^2 \: = \: | {\mathbb Z}_4 |^2
\end{equation}
copies of $[X/{\mathbb Z}_2]$, as expected.

\subsection{Second nonbanded example}

Next we consider $[X/A_4]$, where $A_4$ is the 12-element
alternating group on four indeterminates, with trivially-acting
normal subgroup ${\mathbb Z}_2 \times {\mathbb Z}_2$,
as in \cite[section 2.0.5]{Pantev:2005rh}, \cite[section 5.5]{Hellerman:2006zs}.
We can write
\begin{equation}
    {\mathbb Z}_2 \times {\mathbb Z}_2 \: = \: \{1, \alpha,
    \beta, \gamma \},
\end{equation}
where
\begin{equation}
    \alpha = (14)(23), \: \: \:
    \beta = (13)(24), \: \: \:
    \gamma = (12)(34),
\end{equation}
and all the elements of $A_4$, arranged into
$A_4 / {\mathbb Z}_2 \times {\mathbb Z}_2 = {\mathbb Z}_3$ cosets, are
\begin{eqnarray}
    A_4 & = & \{ 1, \alpha, \beta, \gamma,
    \\
    & & \hspace*{0.25in} (123), (142), (243), (134),
    \\
    & & \hspace*{0.25in} (132), (124), (234), (143) \}.
\end{eqnarray}
The conjugacy classes in $A_4$ are
\begin{equation}
    \{ 1 \}, \: \: \: \{ \alpha, \beta, \gamma \}, \: \: \:
    \{ (123), (142), (243), (134) \}, \: \: \:
    \{ (132), (124), (234), (143) \}.
\end{equation}

As before, we enumerate contributions to a $T^2$ partition function from each sector defined by a loop around the third $S^1$ with a trivially-acting group element (representing a 
conjugacy class) inserted.

\begin{itemize}
    \item $1$:  First, we consider the case that the identity is inserted along the third $S^1$.  Then, we simply count commuting pairs in $A_4$, which are
    \begin{equation}
        {\scriptstyle 1, \alpha, \beta, \gamma} \square_{1, \alpha, \beta, \gamma}, \: \: \:
        {\scriptstyle 1} \square_{(123), \cdots } ,
        %(142), (243), (134)}, 
        \: \: \:
        {\scriptstyle 1} \square_{(132), 
        \cdots } %(124), (234), (143)}, 
        \: \: \:
        {\scriptstyle (123), \cdots } 
        %(142), (243), (134) }
        \square_1, \: \: \:
        {\scriptstyle (132), \cdots }
        %(124), (234), (143)} 
        \square_1,% \: \: \:
    \end{equation}
    \begin{equation}
        {\scriptstyle (123), (132)} \square_{(123), (132)}, \: \: \:
        {\scriptstyle (142), (124)} \square_{(142), (124)}, \: \: \:
        {\scriptstyle (243), (234)} \square_{(243), (234)}, \: \: \:
        {\scriptstyle (134), (143)} \square_{(134), (143)}
    \end{equation}
    which contribute
    \begin{eqnarray}
    \lefteqn{
        \frac{1}{| A_4 |} \bigl[ (16) Z_{1,1} + (4) Z_{1,\xi} + (4) Z_{1,\xi^2} + (4) Z_{\xi,1} + (4) Z_{\xi^2,1} 
        }
     \nonumber \\
     &\hspace*{0.25in} & 
     + (4) Z_{\xi,\xi} + (4) Z_{\xi,\xi^2} + (4) Z_{\xi^2,\xi} + (4) Z_{\xi^2,\xi^2} \bigr]
     \\
     & \hspace*{0.25in}  = & Z\left( X \coprod [X/{\mathbb Z}_3] \right),
    \end{eqnarray}
    where $\xi$ denotes the generator of the effectively-acting ${\mathbb Z}_3 = A_4 / {\mathbb Z}_2 \times {\mathbb Z}_2$,
    as expected from decomposition in the corresponding two-dimensional orbifold \cite[section 5.5]{Hellerman:2006zs}.
    \item $\{ \alpha, \beta, \gamma\}$: In this sector, the conjugacy class represented by $\alpha$, the commuting pairs which also commute with $\alpha$ are
    \begin{equation}
        {\scriptstyle 1, \alpha, \beta, \gamma} \square_{1, \alpha, \beta, \gamma}, 
    \end{equation}
    which contribute
    \begin{equation}
        \frac{1}{| C(\alpha) | } \left[ (16) Z_{1,1} \right] \: = \: (4) Z(X).
    \end{equation}
\end{itemize}
Putting this together, the sum of the results for the
$\{ 1 \}, \{\alpha, \beta, \gamma \}$ sectors are
consistent with a decomposition into
\begin{equation}
    \left( \coprod_5 X \right)  \coprod [X/{\mathbb Z}_3].
\end{equation}
As observed earlier, this reproduces the decomposition prediction of
section~\ref{sect:predict}.

As a consistency check, note that if we interpret
$X$ as a triple cover of $[X/{\mathbb Z}_3]$, then locally this is 
\begin{equation}
    (5)(3) + 1 \: = \: 16 \: = \: | {\mathbb Z}_2 \times {\mathbb Z}_2 |^2
\end{equation}
copies of $[X/{\mathbb Z}_3]$.

\subsection{Third nonbanded example}

Next, consider the orbifold $[X/D_n]$, with trivially-acting
${\mathbb Z}_n \subset D_n$, as in \cite[section 5.6]{Hellerman:2006zs}.  Here, $D_n$ denotes the $2n$-element dihedral group, generated by $a, b$, where
\begin{equation}
    a^2 = 1, \: \: \: b^n = 1, \: \: \: a b a = b^{-1}.
\end{equation}
The trivially-acting subgroup ${\mathbb Z}_n = \langle b \rangle$, and $D_n / {\mathbb Z}_n = {\mathbb Z}_2$.  In the special case that $n$ is even, $D_n$ has a ${\mathbb Z}_2$ center, generated by $z = b^{n/2}$.
For completeness, the elements of $D_n$ can be enumerated as
\begin{equation}
    D_n \: = \: \left\{ 1, b, b^2, \cdots, b^{n-1}, a,
    ab, \cdots, ab^{n-1} \right\}.
\end{equation}
If $n$ is even, then $D_n$ has 
\begin{equation}
    \frac{n}{2} + 3
\end{equation}
conjugacy classes, which are, explicitly
\begin{equation}
    \{ 1 \}, \: \: \:
    \{ a, ab^2, ab^4, \cdots, ab^{n-2} \}, \: \: \:
    \{ ab, ab^3, ab^5, \cdots, ab^{n-1} \},
\end{equation}
\begin{equation}
    \{ b^j, b^{-j} \} \: \: \: \mbox{ for } 1 \leq j \leq n/2.
\end{equation}
If $n$ is odd, then $D_n$ has
\begin{equation}
    \frac{n+3}{2}
\end{equation}
conjugacy classes, which are, explicitly
\begin{equation}
    \{ 1 \}, \: \: \: 
    \{a, ab, ab^2, ab^3, ab^4, \cdots, ab^{n-1} \},
\end{equation}
\begin{equation}
    \{ b^j, b^{-j} \} \: \: \: \mbox{ for } 1 \leq j \leq \frac{n-1}{2}.
\end{equation}

As before, we consider contributions to a $T^2$ partition function from sectors with trivially-acting conjugacy classes on the third $S^1$.
\begin{itemize}
    \item $1$:  In this sector, all commuting pairs in $D_n$ contribute.  For $n$ odd, these are
    \begin{equation}
        {\scriptstyle \langle b \rangle} \square_{\langle b \rangle}, \: \: \:
        {\scriptstyle 1} \square_{a, ab, \cdots}, \: \: \:
        {\scriptstyle a, ab, \cdots} \square_1, \: \: \:
        {\scriptstyle a b^i} \square_{a b^i}
    \end{equation}
    which contribute
    \begin{eqnarray}
    \lefteqn{
        \frac{1}{| D_n |} \left[ (n^2) Z_{1,1} + (n) Z_{1,\xi} + (n) Z_{\xi,1} + (n) Z_{\xi,xi} \right]
        }
        \\
        & = & \frac{1}{2n} \left[ (n^2-n) Z_{1,1} + (n)\left( Z_{1,1} + Z_{1,\xi} + Z_{\xi,1} + Z_{\xi,\xi} \right) \right],
        \\
        & = & Z\left(  \coprod_{(n-1)/2} X \coprod_1 [X/{\mathbb Z}_2] \right), 
    \end{eqnarray}
    where $\xi$ generates the effectively-acting
    ${\mathbb Z}_2 = D_n/{\mathbb Z}_n$.

    For $n$ even, in addition to the commuting pairs above,
    there are also
    \begin{equation}
        {\scriptstyle z} \square_{a, ab, \cdots}, \: \: \:
        {\scriptstyle a, ab, \cdots} \square_z, \: \: \:
        {\scriptstyle a b^{i+n/2}} \square_{a b^i},
    \end{equation}
    and the total contribution becomes
    \begin{eqnarray}
    \lefteqn{
        \frac{1}{| D_n |} \left[ (n^2) Z_{1,1} + (2n) Z_{1,\xi} + (2n) Z_{\xi,1} + (2n) Z_{\xi,xi} \right]
        }
        \\
        & = & \frac{1}{2n} \left[ (n^2-2n) Z_{1,1} + (2n) \left( Z_{1,1} + Z_{1,\xi} + Z_{\xi,1} + Z_{\xi,\xi} \right) \right],
        \\
        & = & Z\left( \coprod_{(n-2)/2} X \, \coprod_2 [X/{\mathbb Z}_2 ] \right).
    \end{eqnarray}
    \item $z$:  If $n$ is even, then $z = b^{n/2}$ is in the center of $D_n$ and defines its own conjugacy class.  In this sector, there are the same
    contributions as for the $1$ sector above, and so we get the contribution
    \begin{equation}
        Z\left( \coprod_{(n-2)/2} X \, \coprod_2 [X/{\mathbb Z}_2 ] \right).
    \end{equation}
    \item $\{ b^i, b^{-i}\}$, $i \neq 0, n/2$:  In these sectors, there are contributions from the commuting pairs
    \begin{equation}
        {\scriptstyle \langle b \rangle} \square_{ \langle b \rangle},
    \end{equation}
    which contribute
    \begin{equation}
        \frac{1}{|C(b^i)|} \left( n^2 Z_{1,1} \right) \: = \:
        n Z(X) \: = \: Z\left( \coprod_n X \right).
    \end{equation}
\end{itemize}

We summarize the total contribution as follows.
If $n$ is odd, the results are consistent with a decomposition into
\begin{equation}
    \coprod_{ (n+1)(n-1)/2 } X \, \coprod_1 [X/{\mathbb Z}_2].
\end{equation}
As observed earlier, this reproduces the decomposition prediction of
section~\ref{sect:predict}.

As a consistency check, if we interpret $X$ as a double cover of $[X/{\mathbb Z}_2]$, then this is locally
\begin{equation}
    (2) \frac{(n+1)(n-1)}{2} + 1 \: = \: n^2 \: = \: | {\mathbb Z}_n |^2
\end{equation}
copies of $[X/{\mathbb Z}_2]$.

If $n$ is even, the results are consistent with a decomposition into
\begin{equation}
    \prod_{(n-2)(n+2)/2} X \, \coprod_4 [X/{\mathbb Z}_2].
\end{equation}

As a consistency check, if we interpret $X$ as a double cover of $[X/{\mathbb Z}_2]$, then this is locally
\begin{equation}
    (2) \frac{(n-2)(n+2)}{2} + 4 \: = \: n^2 \: = \: | {\mathbb Z}_n |^2
\end{equation}
copies of $[X/{\mathbb Z}_2]$.

As observed earlier, for $n$ both even and odd,
this reproduces the decomposition prediction of
section~\ref{sect:predict}.

\subsection{First trivially-acting nonabelian group example}

Next, consider the orbifold $[X/S_3]$, where all of $S_3$ acts trivially.
We enumerate the elements of $S_3$ as
\begin{equation}
    S_3 \: = \: \{ 1, (12), (23), (13), (123), (132) = (123)^2 \}.
\end{equation}
The conjugacy classes are 
\begin{equation}
    \{ 1 \}, \: \: \: \{ (12), (23), (13) \}, \: \: \:
    \{ (123), (132) \}.
\end{equation}

As before, we itemize contributions to a $T^2$ partition function by sectors corresponding to conjugacy classes of trivially-acting group elements on the third $S^1$.  For each conjugacy class, we pick one representative below.
\begin{itemize}
    \item $1$:  In this sector, $C(1) = S_3$, so all commuting pairs in $S_3$ contribute.  These are
    \begin{equation}
        {\scriptstyle 1} \square_1, \: \: \:
        {\scriptstyle 1} \square_{ g \neq 1}, \: \: \:
        {\scriptstyle g \neq 1} \square_1, \: \: \:
        {\scriptstyle (12)} \square_{(12)}, \: \: \:
        {\scriptstyle (23)} \square_{(23)}, \: \: \:
        {\scriptstyle (13)} \square_{(13)}, \: \: \:
        {\scriptstyle (123), (132)} \square_{(123), (132)}
    \end{equation}
    which contribute
    \begin{equation}
        \frac{1}{|S_3|} \left[ 1 + 5 + 5 + 1 + 1 + 1 + 4 \right] Z_{1,1} \: = \: (3) Z(X),
    \end{equation}
    consistent with the prediction of decomposition in the corresponding two-dimensional orbifold
    (as $S_3$ has three irreducible representations).
    \item $(12)$:  In this sector, $C( (12) ) = \{ 1, (12) \}$, so the pertinent commuting pairs are
    \begin{equation}
        {\scriptstyle 1, (12)} \square_{ 1, (12)}
    \end{equation}
    which contribute
    \begin{equation}
        \frac{1}{|C( (12) ) |} \left(4 \right) Z_{1,1} \: = \: (2) Z(X).
    \end{equation}
    \item $\{ (123), (132) \}$:  In this sector, $C( (123) ) = \{ 1, (123), (132) \}$,
    and the pertinent commuting pairs are
    \begin{equation}
        {\scriptstyle 1, (123), (132)} \square_{1, (123), (132)}
    \end{equation}
    which contribute
    \begin{equation}
        \frac{1}{| C( (123) )|} \left( 9 \right) Z_{1,1} \: = \: (3) Z(X).
    \end{equation}
\end{itemize}

Altogether, adding the contributions from the different sectors gives
$(8) Z(X)$, 
which is consistent with a decomposition into
\begin{equation}
    \coprod_8 X.
\end{equation}
As observed earlier, this reproduces the decomposition prediction of
section~\ref{sect:predict}.

\subsection{Second trivially-acting nonabelian group example}

Next, consider the orbifold $[X/{\mathbb H}]$, where all of ${\mathbb H} = \{\pm 1, \pm i, \pm j,
\pm k\}$ acts trivially.  The conjugacy classes are
\begin{equation}
    \{ 1 \}, \: \: \: \{ -1 \}, \: \: \:
    \{ \pm i \}, \: \: \: \{ \pm j \}, \: \: \:
    \{ \pm k \}.
\end{equation}

As before, we itemize contributions to a $T^2$ partition function by sectors corresponding to group elements representing trivially-acting conjugacy classes on the third $S^1$:
\begin{itemize}
    \item $\pm 1$:  In these two sectors, all commuting pairs in ${\mathbb H}$ contribute.  These are
    \begin{equation}
        {\scriptstyle \pm 1} \square_{\pm 1}, \: \: \:
        {\scriptstyle \pm 1} \square_{g \neq \pm 1}, \: \: \:
        {\scriptstyle g \neq \pm 1} \square_{\pm 1}, \: \: \:
        {\scriptstyle \pm i} \square_{\pm 1}, \: \: \:
        {\scriptstyle \pm j} \square_{\pm j}, \: \: \:
        {\scriptstyle \pm k} \square_{\pm k},
    \end{equation}
    which contribute
    \begin{equation}
        \frac{1}{| {\mathbb H}| }\left( 4 + (2)(6) + (2)(6) + 4 + 4 + 4\right) Z_{1,1} \: = \:
        (5) Z(X),
    \end{equation}
    consistent with the prediction of decomposition in the corresponding two-dimensional orbifold (as ${\mathbb H}$ has five irreducible representations).
    \item $\{ \pm i \}$:  In this sector, the pertinent commuting pairs are
    \begin{equation}
        {\scriptstyle \pm 1, \pm i} \square_{ \pm 1, \pm i},
    \end{equation}
    which contribute
    \begin{equation}
        \frac{1}{| C(i)|} (4^2) Z_{1,1} \: = \: \frac{16}{4} Z(X) \: = \: (4) Z(X).
    \end{equation}
    \item $\{ \pm j\}$, $\{\pm k\}$:  These two sectors each give the same results as the $\{ \pm i \}$ sector.
\end{itemize}
Altogether, adding the contributions from the different sectors gives $$(22) Z(X),$$
which is consistent with a decomposition into
\begin{equation}
    \coprod_{22} X.
\end{equation}
As observed earlier, this reproduces the decomposition prediction of
section~\ref{sect:predict}.

\subsection{Third trivially-acting nonabelian group example}

Next, consider the orbifold $[X/D_4]$, where all of the eight-element dihedral group $D_4$ acts trivially.
The conjugacy classes of $D_4$ are
\begin{equation}
    \{ 1 \}, \: \: \: \{ z \}, \: \: \:
    \{a, az \}, \: \: \:
    \{ b, bz \}, \: \: \:
    \{ ab, ba \}.
\end{equation}

As before, we itemize contributions to a $T^2$ partition function by sectors corresponding to trivially-acting group elements on the third $S^1$:
\begin{itemize}
    \item $1, z$:  In these two sectors, all commuting pairs in $D_4$ contribute.  These are
    \begin{equation}
        {\scriptstyle 1, z} \square_{1, z}, \: \: \:
        {\scriptstyle 1, z} \square_{g \neq 1, z}, \: \: \:
        {\scriptstyle g \neq 1, z} \square_{1, z}, \: \: \:
        {\scriptstyle a, az} \square_{a, az}, \: \: \:
        {\scriptstyle b, bz} \square_{b, bz}, \: \: \:
        {\scriptstyle ab, ba} \square_{ab ba},
    \end{equation}
    which contribute
    \begin{equation}
        \frac{1}{| D_4|} \left[ 4 + (2)(6) + (2)(6) + 4 + 4 + 4\right] Z_{1,1}
        \: = \: (5) Z(X),
    \end{equation}
    consistent with the prediction of decomposition in the corresponding two-dimensional orbifold (as $D_4$ has five irreducible representations).
    \item $\{ a, az\}$: In this sector, the pertinent commuting pairs are
    \begin{equation}
        {\scriptstyle 1, z} \square_{1, z}, \: \: \:
        {\scriptstyle 1, z} \square_{a, az}, \: \: \:
        {\scriptstyle a, az} \square_{1, z}, \: \: \:
        {\scriptstyle a, az} \square_{a, az},
    \end{equation}
    which contribute
    \begin{equation}
        \frac{1}{ | C(a) | } \left[ 4 + 4 + 4 + 4 \right] Z_{1,1} \: = \:
        (4) Z(X).
    \end{equation}
    \item $\{b, bz\}, \{ ab, ba\}$:  Contributions from each of these two sectors follow the same form as those from the $\{a, az\}$ sector.
\end{itemize}

Altogether, adding the contributions from the different sectors gives
$(22) Z(X)$, which is consistent with a decomposition into
\begin{equation}
    \coprod_{22} X.
\end{equation}
As observed earlier, this reproduces the decomposition prediction of
section~\ref{sect:predict}.

\section{Quantum K theory via GLSM computations}   \label{sect:glsm}

In this section, we turn to a different class of presentations, namely gauged linear
sigma models, rather than orbifolds, and verify the structure predicted in
section~\ref{sect:predict}.  Here, the trivially-acting subgroup $K$ of the gauge group will always be a subset of the center.  
Furthermore, in all subsections except \ref{sect:glsm:badlevel},
the two-dimensional theory will have both a $BK$ (one-form) and $K$ (zero-form) symmetry, without
a 't Hooft anomaly between them.
As a result, in all subsections except \ref{sect:glsm:badlevel},  
we will see a decomposition of the effective two-dimensional theory
into $|K|$ copies of a two-dimensional GLSM with a one-form symmetry, each copy of which again separately decomposes,
giving altogether a decomposition into a total of $|K|^2$ universes, as originally predicted for such cases in
\cite{Gu:2021yek,Gu:2021beo}.  
The first level of decomposition -- the choice of Wilson line -- is visible in the fact that
the roots are symmetric under multiplication by $\ell$th root of unity.
As the Coulomb branch parameters are the Wilson lines, this is precisely a symmetry between
possible choices of Wilson line.
Furthermore, by using GLSM methods,
we are able to make predictions for the quantum K theory rings, not just the structure
of the decomposition.

In section~\ref{sect:glsm:badlevel}, we consider more general Chern-Simons levels and 't Hooft anomalies, and discuss how the decomposition story is modified.

We should emphasize that, as is typical in such computations, we are making mathematical conjectures via an interpretation of the Coulomb branch equations, and a proper treatment of the mathematics should await a more rigorous mathematical analysis of the quantum K theory.  Our point, however, is that an interpretation of the Coulomb branch equations consistent with expectations from decomposition is possible.

\subsection{Warmup: Gerbes on projective spaces}
\label{sect:proj}

\subsubsection{General ${\mathbb Z}_{\ell}$ gerbes on projective spaces}

We begin with general ${\mathbb Z}_{\ell}$ gerbes on
${\mathbb P}^{n-1}$, reviewing examples discussed in \cite[section 4]{Gu:2021yek}.  Following e.g.~\cite{Pantev:2005zs},
such gerbes can be described by a $U(1)^2$ GLSM with fields $x_i$, $z$, and charges
\begin{center}
\begin{tabular}{cr}
$x_i$ & $z$ \\ \hline
$1$ & $-m$ \\
$0$ & $\ell$
\end{tabular}
\end{center}
As discussed in \cite{Pantev:2005zs}, this defines a
${\mathbb Z}_{\ell}$ gerbe over ${\mathbb P}^{n-1}$,
with characteristic class $-m \mod \ell$.  The weighted projective
space ${\mathbb P}^{n-1}_{[\ell,\cdots,\ell]}$ is equivalent to the
case $m=1$.

Before computing the quantum K theory, we begin by reviewing the quantum cohomology.
Following \cite[section 3.2]{Pantev:2005zs}, the quantum cohomology ring computed from the two-dimensional GLSM above is
\begin{equation}
    {\mathbb C}[\psi_1,\psi_2] / \langle \psi_1^n - \psi_2^m q_1, \psi_2^{\ell} - q_2 \rangle.
\end{equation}
(This is computed using GLSM Coulomb branch methods, as the critical locus of the one-loop twisted effective
superpotential given in e.g.~\cite{Morrison:1994fr}.)
In decomposition, we interpret the $\psi_2$ as indexing $\ell$ different universes, each a copy of 
the ordinary supersymmetric ${\mathbb P}^{n-1}$ model with $q$'s in different universes slightly different, related by shifts of the $B$ field.

Let us next compute the quantum K theory of these examples.
Physically, this is the OPE ring of Wilson lines in a three-dimensional gauge theory on a 3-manifold of the
form $S^1 \times \Sigma$, for $\Sigma$ a Riemann surface,
where the Wilson lines are wrapped on $S^1$ and move in parallel along $\Sigma$ \cite{Bullimore:2014awa,Jockers:2018sfl,Jockers:2019wjh,Jockers:2019lwe}.
This also can be computed from the critical locus of a twisted one-loop effective superpotential along
the Coulomb branch, albeit in a two-dimensional gauge theory arising from a regularized Kaluza-Klein reduction of a three-dimensional
gauge theory \cite{Ueda:2019qhg,Nekrasov:2009uh,Closset:2016arn,Closset:2018ghr,Closset:2019hyt,Closset:2023vos,Closset:2023bdr,Gu:2020zpg,Gu:2022yvj,Gu:2023tcv,Gu:2023fpw}.

The two-dimensional theory has an infinite tower of fields.  For each field in the three-dimensional theory, transforming in representation $R$ of the gauge group
$G$, there is an infinite tower of massive fields in the two-dimensional theory, each also transforming in the same representation $R$ of the gauge group.
As a result, if the three-dimensional gauge theory has a one-form symmetry due to some subgroup of the gauge group acting trivially on all the matter, the two-dimensional theory arising from the Kaluza-Klein reduction will have the same one-form symmetry, a fact we shall use throughout this paper.

The twisted one-loop effective superpotential arising from those Kaluza-Klein towers is then regularized.
Following \cite[equ'n (2.1)]{Gu:2020zpg}, \cite[equ'n (2.33)]{Closset:2016arn}, \cite[section 2.2.2]{Nekrasov:2009uh},
\cite[section 2.2.1]{Closset:2017zgf}, \cite[section 2]{Closset:2019hyt}, \cite[equ'n (2.10)]{Closset:2023vos}, 
\cite[equ'n (2.2)]{Closset:2023bdr},
the regularized superpotential has the form
\begin{equation}
W \: = \: \frac{1}{2} k^{ab} (\ln X_a) (\ln X_b) \: + \:
\sum_a (\ln q_a) (\ln X_a)
\: + \:
\sum_i\left[ 
{\rm Li}_2\left( X^{\rho_i} \right) \: + \:
\frac{1}{4} \left( \ln\left( X^{\rho_i} \right) \right)^2 \right],
\end{equation}
where
\begin{equation}
X^{\rho_i} \: = \: \prod_a X_a^{Q^i_a},
\end{equation}
so that
\begin{equation}
X^x \: = \: X_1, \: \: \:
X^z \: = \: X_1^{-m} X_2^{\ell},
\end{equation}
and
\begin{equation}
k^{11} \: = \: -\frac{1}{2} \left( n + m^2 \right),
\: \: \:
k^{12} = k^{21} \: = \: \frac{1}{2} m \ell,
\: \: \:
k^{22} \: = \: -\frac{1}{2} \ell^2,
\end{equation}
using the general formula for Chern-Simons levels \cite[equ'n (2.6)]{Gu:2023tcv} 
\begin{equation}
k^{ab} \: = \: \frac{1}{2} \sum_i \left( R_i - 1 \right)
Q_i^a Q_i^b,
\end{equation}
as given by $U(1)_{-1/2}$ quantization, see e.g.~\cite[section 2.2]{Closset:2019hyt}, \cite{Gu:2020zpg},
as that is the choice that reproduces ordinary quantum K-theory.

Computing the critical locus, we find the quantum K theory ring relations
\begin{eqnarray}
(1-X_1)^{n} \left( 1 - X_1^{-m} X_2^{\ell} \right)^{-m}
& = & q_1,
\\
\left( 1 - X_1^{-m} X_2^{\ell} \right)^{\ell} & = & q_2.
\end{eqnarray}

In passing, note that shifting $m \mapsto m + \ell$ is equivalent to
changing $q_1 \mapsto q_1 q_2^{-1}$. It is in this sense that the quantum K theory
ring only depends upon the characteristic class of the gerbe, $m \mod \ell$.

Define
\begin{equation}
    y \: = \:  1 - X_1^{-m} X_2^{\ell} ,
\end{equation}
then we can write
the ring relations above as
\begin{eqnarray}
    \left( 1 - X_1 \right)^{n} & = & q_1 y^m,    \label{eq:qk:ggerbe:pn-1:1}
    \\
    y^{\ell} & = & q_2.
\end{eqnarray}

As $y$ is determined by $X_2^{\ell}$, there is an $\ell$-fold phase choice in solutions of $X_2$ for fixed $y$, which we interpret as
reflecting $\ell$ copies of a theory, each copy of which itself decomposes into $\ell$ universes,
for altogether 
$\ell^2$ copies of the quantum K theory ring
of the underlying space ${\mathbb P}^{n-1}$ (from the first relation), indexed by the $\ell^2$ values of $y$ and $X_2$.

\subsubsection{Special case: Weighted projective spaces}

Next, we specialize to gerbes which can be presented as weighted projective spaces, a case previously discussed in \cite[section 4.1]{Gu:2021yek}.
These admit an additional, different, UV presentation, so it will be instructive to compute
the quantum K theory ring and compare to the general case, reviewing the result of \cite[section 4.1]{Gu:2021yek}.

For a general stacky weighted projective space ${\mathbb P}^{n-1}_{[w_0,\cdots,w_{n-1}]}$,
following the prescription in e.g.~\cite[section 2]{Gu:2020zpg},
the quantum K theory ring relations arise as the critical locus of the
superpotential
\begin{equation}
W \: = \: \frac{k}{2} \left( \ln X \right)^2 \: + \:
(\ln q) (\ln X) \: + \: \sum_i \left[
{\rm Li}_2\left( X^{w_i} \right) \: + \: 
\frac{1}{4} \left( \ln\left( X^{w_i} \right) \right)^2 \right],
\end{equation}
with Chern-Simons level
\begin{equation}
k \: = \: - \frac{1}{2} \sum_i (w_i)^2.
\end{equation}

It is straightforward to compute that the critical locus is given by
\begin{equation}
\prod_{j=0}^{n-1} \left( 1 - X^{w_j} \right)^{w_j} \: = \: q.
\end{equation}
This result matches results for
quantum K theory rings for weighted projective stacks given in
\cite[thm.~1.5, cor.~5.6]{zhangstacks}, namely that for a weighted
projective space ${\mathbb P}^{n-1}_{[w_0,\cdots,w_{n-1}]}$.  (See also
\cite[example 1.3]{gw}, and \cite{Gu:2021yek} for a physics discussion.)

Now, ${\mathbb Z}_{\ell}$ gerbes on ${\mathbb P}^{n-1}$ of characteristic class $-1 \mod \ell$ are described as
stacky weighted projective spaces ${\mathbb P}^{n-1}_{[\ell,\ell,\cdots,\ell]}$,
meaning that all weights $w_i$ are equal to $\ell$.  (For a discussion of how the resulting
physical theory in two dimensions differs from that of the ordinary supersymmetric
${\mathbb P}^{n-1}$ model, see \cite{Pantev:2005rh,Pantev:2005wj,Pantev:2005zs}.)

In the case that the weights are all $\ell$,
the result above for the quantum K theory ring reduces to
\begin{equation}  \label{eq:qk:pn-1}
    (1 - X^{\ell})^{\ell n} = q,
\end{equation}

As a consistency check, write $X = \exp(-2 \pi R \sigma)$,
$q = R^{\ell n} q_{2d}$, then the relation~(\ref{eq:qk:pn-1}) reduces
to
\begin{equation}
\sigma^{\ell n} \: \propto \: q_{2d},
\end{equation}
which is the same quantum cohomology ring relation for
${\mathbb P}^{n-1}_{[\ell,\cdots,\ell]}$ discussed in
e.g.~\cite{Pantev:2005rh,Pantev:2005wj,Pantev:2005zs}.

To help explain~(\ref{eq:qk:pn-1}), consider the case that $\ell=1$,
describing an ordinary projective space ${\mathbb P}^{n-1}$.
In this case, we interpret $X = {\cal O}(-1)$, then $1-X = {\cal O}_H$,
the class of a hyperplane.  The product corresponds to generic intersection,
and the intersection of $n$ hyperplanes in general position in
${\mathbb P}^{n-1}$ is empty, so that classically,
$(1-X)^n = 0$ in $K({\mathbb P}^{n-1})$.

In the case $\ell > 1$, our understanding of \cite{zhangstacks,jkk,arz} is that to interpret~(\ref{eq:qk:pn-1}),
one should interpret $X = {\cal O}(-1/\ell)$, a line bundle on the gerbe which is not a pullback from the underlying projective space, and that one uses an orbifold product on the inertia stack, a K-theoretic analogue of products in
orbifold cohomology.

In terms of decomposition, taking an $\ell$th root of~(\ref{eq:qk:pn-1}), we get $\ell$ copies of the ordinary
quantum K theory ring of ${\mathbb P}^{n-1}$, generated by $X^{\ell}$, indexed by $\ell$th roots of unity.  The fact that the generator is $X^{\ell}$ can be interpreted as meaning we get the decomposition above for each of the $\ell$ roots of $X^{\ell}$, giving a decomposition into a total of $\ell^2$ universes, matching expectations.
As a consistency check, we can 
Compare to the ring relations for general ${\mathbb Z}_{\ell}$ gerbes as follows.
Identify $q = q_1^{\ell} q_2^{m}$ and $X_1 = X^{\ell}$, 
then the $\ell$th power of the relation~(\ref{eq:qk:ggerbe:pn-1:1}) can be written
\begin{equation}
\left( 1 - X^{\ell} \right)^{\ell n} \: = \: q.
\end{equation}
which, for generator $X^{\ell}$, is the ring relation for $m=1$, matching~(\ref{eq:qk:pn-1}).

In any event, again we see a decomposition in quantum K theory, as predicted.

\subsection{General ${\mathbb Z}_{\ell}$ gerbes over Grassmannians}   \label{sect:ggerbe:gkn}

In this section we will discuss ${\mathbb Z}_{\ell}$ gerbes over Grassmannians,
constructing them as ${\mathbb C}^{\times}$ quotients of line bundles over
ordinary Grassmannians, exactly as we reviewed for projective spaces in section~\ref{sect:proj}.

It will be handy to recall that 
irreducible representations of $u(k)$ are characterized by a $k$-tuple of ordered integers
\begin{equation}
    [\lambda_1, \lambda_2, \dots , \lambda_k], \quad \lambda_i \in \mathbb{Z}
\end{equation}
with $\lambda_1 \ge \lambda_2 \ge \dots \ge \lambda_k$.
In this description, the corresponding
$su(k)$ representation is
\begin{equation}
    (\lambda_1 - \lambda_k, \lambda_2 - \lambda_k, \dots \lambda_{k-1}-\lambda_k),
\end{equation}
where the $\det u(k) = u(1)$ charge is given by $\lambda_1 + \lambda_2 + \dots + \lambda_k$.

\subsubsection{Quick review of ordinary Grassmannians}

Before describing gerbes on Grassmannians, and their quantum cohomology and quantum K theory,
let us first quickly review ordinary Grassmannians, their quantum cohomology and quantum K theory.

First, an ordinary Grassmannian $G(k,n)$ can be constructed as ${\mathbb C}^{kn}//GL(k)$, meaning a $U(k)$ GLSM
with $n$ chiral multiplets in the fundamental representation \cite{Witten:1993xi}.  As a Fano space realized by a GLSM, both its quantum cohomology and quantum K theory can be computed using Coulomb branch methods.

Recall that for an ordinary Grassmannian $G(k,n)$, the nonequivariant relations in the quantum cohomology ring are
\begin{equation}  
    (-1)^{k-1} q = \sigma_a^n,
\end{equation}
where $\{ \sigma_a \}$ ($a \in \{1, \cdots, k\}$)
are local Coulomb branch coordinates on a Weyl-group orbifold, or more formally,
Chern roots of the universal subbundle $S$.
After symmetrization, these
give rise to the relation
\begin{equation}
    c(S) \, c(Q) = 1 + (-1)^{n-k} q.
\end{equation}
This is the quantum cohomology ring relation (see e.g.~\cite{Witten:1993xi}).

There is a similar description\footnote{
Quantum K theory of Grassmannians has been extensively discussed,
see e.g.~\cite{Ueda:2019qhg,Jockers:2019lwe,Gu:2020zpg,Closset:2023bdr,Givental:2001clq,Koroteev:2017nab,Givental:1993nc,Kim:1996}
for a few examples in both the mathematics and physics literatures.
} of the quantum K theory ring of a Grassmannian $G(k,n)$,
derived again from a twisted one-loop effective superpotential.
The Coulomb branch equations one derives are
(see e.g.~\cite[equ'n (2.40)]{Gu:2020zpg})
\begin{equation}  \label{eq:basicgkn:qk}
    (-1)^{k-1} q X_a^k = (\det X) (1-X_a)^n,
\end{equation}
(where the $X_a = \exp(-2\pi R \sigma_a)$ for $R$ the radius of the $S^1$ in the 3-manifold,)
which after symmetrization become the quantum K-theory ring relation \cite[theorem 1.1]{Gu:2022yvj}
\begin{equation}
    \lambda_y(S) \star \lambda_y( {\mathbb C}^n/S) \: = \:
    \lambda_y( {\mathbb C}^n) \: - \: y^{n-k} \frac{q}{1-q} \det( {\mathbb C}^n/S) \star \left( \lambda_y(S) - 1 \right).
\end{equation}
For the purposes of comparing to decomposition predictions, it will mostly be sufficient for our purposes to work only with Coulomb branch expressions.

\subsubsection{Description of ${\mathbb Z}_{\ell}$ gerbes over Grassmannians}

Mathematically, we can describe ${\mathbb Z}_{\ell}$ gerbes over Grassmannians
$G(k,n)$ as quotients
\begin{equation}
    \left[ L^* / {\mathbb C}^* \right],
\end{equation}
where $L \rightarrow G(k,n)$ is a line bundle,
$L^*$ is $L$ minus its zero section, and
${\mathbb C}^*$ acts with a trivially-acting
${\mathbb Z}_{\ell}$ subgroup.
The resulting ${\mathbb Z}_{\ell}$ gerbe has characteristic class $c_1(L) \mod \ell$.

In physics, we describe such gerbes as\footnote{
We would like to thank W.~Gu for useful discussions.
} a $U(k) \times U(1)$ gauge theory with $n$ chiral superfields in the fundamental representation of $U(k)$ and neutral under $U(1)$, together with another chiral superfield in the $U(k)$ representation $[-m', -m', \dots, -m']$ (equivalently, the charge $-m'k$ representation of $\det U(k)$) and of charge $\ell$ under the $U(1)$.  This describes a ${\mathbb Z}_{\ell}$ gerbe on $G(k,n)$ of characteristic class $-k m' \mod \ell$.

\subsubsection{Quantum cohomology} \label{eq:ggerbe:qh}

Using standard methods \cite{Morrison:1994fr}, the quantum cohomology ring of this Grassmannian is given
by
\begin{eqnarray}
\sigma_a^n \left( \ell \sigma_0 - m' \sum_b \sigma_b \right)^{-m'}
& = & (-)^{k-1} \tilde{q}_1,
\\
\left(  \ell \sigma_0 - m' \sum_b \sigma_b \right)^{\ell} & = & \tilde{q}_0.
\end{eqnarray}

Define
\begin{equation}
\Upsilon \: = \:  \ell \sigma_0 - m' \sum_b \sigma_b.
\end{equation}
Rescaling $\tilde{q}_0 \rightarrow 1$ without\footnote{
It can be absorbed into $q_1$ with suitable redefinitions.}
loss of generality, we can write these equations as
\begin{equation}
\sigma_a^n \: = \: (-)^{k-1} \tilde{q}_1 \Upsilon^{m'},
\: \: \:
\Upsilon^{\ell} = 1.
\end{equation}
This is precisely as expected for the quantum cohomology ring of
a disjoint union of $\ell$ copies of $G(k,n)$, each with $B$ fields / theta
angles slightly shifted (as encoded in $\ell$th roots of unity),
the same pattern discussed in \cite{Pantev:2005zs} for quantum cohomology
rings of toric gerbes on projective spaces.
(For completeness, we also mention that the same structure is visible in
nonabelian mirror constructions \cite{Gu:2018fpm}.)

\subsubsection{Quantum K theory}

Next, we compute the quantum K theory ring relations. Using the results reviewed in section~\ref{sect:proj}, we can write the effective twisted superpotential as follows,
\begin{align}
    W ={}& \frac{k}{2} \sum_{a=1}^k(\ln X_a)^2 - \frac{1}{2} \left(\sum_{a=1}^k \ln X_a\right)^2 + (\ln (-1)^{k-1}q_1) \sum_{a=1}^k \ln X_a \cr
    & + (\ln q_0) (\ln X_0) + n \sum_{a=1}^k \mathrm{Li}_2(X_a) + \mathrm{Li}_2((\det X)^{-m'} X_0^\ell).
\end{align}
Taking the critical locus,
\begin{equation}
    X_0 \frac{\partial W}{\partial X_0} = 0, \quad X_a \frac{\partial W}{\partial X_a} = 0,
\end{equation}
we obtain,
\begin{align}  
    q_0 &= \left(1 - (\det X)^{-m'} X_0^\ell\right)^\ell,
    \label{eq:genlgerbe:1}
    \\
    (-1)^{k-1} q_1 X_a^k &= (\det X) (1-X_a)^n \left(1-(\det X)^{-m'} X_0^\ell\right)^{-m'}.
    \label{eq:genlgerbe:2}
\end{align}
Furthermore, this implies
\begin{equation}\label{eqn:rel-l-power}
    (-1)^{\ell(k-1)} q_0^{m'}q_1^\ell X_a^{k \ell} = (\det X)^\ell (1-X_a)^{\ell n}.
\end{equation}

As a consistency check, note that when $\ell = 1$, we obtain
\begin{equation}
    (-1)^{k-1} (q_0^{m'} q_1) X_a^k = (\det X) (1-X_a)^n,
\end{equation}
which agrees with the relations for ordinary Grassmannian~(\ref{eq:basicgkn:qk})
%\begin{equation}
%    (-1)^{k-1} q X_a^k = (\det X) (1-X_a)^n,
%\end{equation}
if we make the identification $q_0^{m'} q_1 = q$.

Now, to compare to the claimed decomposition, define
\begin{equation}
    y \: = \: 1 - (\det X)^{-m'} X_0^\ell.
\end{equation}
Then, the quantum K-theory ring relations~(\ref{eq:genlgerbe:2}) become
\begin{eqnarray}
    q_0 & = & y^{\ell},
    \\
    (-1)^{k-1} q_1 y^{m'} X_a^k & = & (\det X) \left( 1 - X_a \right)^n,
\end{eqnarray}
which, given the $\ell$-fold ambiguity in $X_0$ for fixed $y$, is clearly a total of $\ell^2$ copies of the quantum K theory ring relations of $G(k,n)$, each with shifted $q$,
shifted by an $\ell$-th root of unity, consistent with expectations.

\subsection{Gerbes via weighted Grassmannians}
\label{sect:wgerbe}

In this section we will construct a theory describing a ${\mathbb Z}_{km+1}$ gerbe
over the Grassmannian, as an analogue of a weighted projective space, and analyze physics predictions for its quantum K theory.  In principle this is just a different presentation of one of the
${\mathbb Z}_{\ell}$ gerbes of the previous section, but it will be an instructive check to consider it in detail.

\subsubsection{Construction of the theory}

First, following \cite{Witten:1993xi}, we describe an ordinary Grassmannian $G(k,n)$  
by a $3d$  $\mathcal{N}=2$ $U(k)$ gauge theory with $n$ chiral superfields in the fundamental representation,
meaning the representation with highest weight
\begin{equation}
    [1, \underbrace{0, \cdots, 0}_{k-1}].
\end{equation}

Now, to describe the weighted Grassmannian describing a ${\mathbb Z}_{km+1}$ gerbe\footnote{
We would like to thank W.~Gu for useful discussions.
},
consider a 3d $\mathcal{N}=2$ $U(k)$ gauge theory with $n$ chiral superfields in the $U(k)$ representation with highest weight 
\begin{equation}
    [m+1, \underbrace{m, \dots, m}_{k-1}],
\end{equation}
meaning, a tensor product of the fundamental representation with a charge $m$ representation of $\det U(k)$.
We can understand why this describes a gerbe as follows.  First, 
since the determinant
of $U(k)$ is itself a product of $k$ $U(1)$ factors, namely the diagonal in a $k \times k$ matrix representation of $U(k)$, there is a trivially-acting ${\mathbb Z}_k$ subgroup
of $U(k)$ acting on anything of charge $1$ under $\det U(k)$, defined by diagonal matrices with $k$th roots of unity along the diagonal.  A charge $m$ representation of $\det U(k)$ is therefore invariant under a ${\mathbb Z}_{mk}$ subgroup of $U(k)$.  The chiral multiplets here are in a charge $m$ representation of $U(k)$ tensored with the fundamental representation of $U(k)$, which is
invariant under a ${\mathbb Z}_{km+1}$ subgroup of $U(k)$.
Thus, this gauge theory
has a trivially-acting ${\mathbb Z}_{km+1}$ subgroup, hence the gauge theory has a ${\mathbb Z}_{km+1}$ one-form
symmetry,
and so
describes a ${\mathbb Z}_{km+1}$ gerbe over $G(k,n)$.

\subsubsection{Quantum cohomology}   \label{sect:wgkn:2d}

Using standard methods \cite{Morrison:1994fr},
it is straightforward to compute that the quantum cohomology ring is given by
\begin{equation}   \label{eq:wgr:qh}
    (-1)^{k-1} q = \left(\sigma_a + m \sum_b \sigma_b\right)^n \prod_{b=1}^k \left(\sigma_b + m \sum_c \sigma_c\right)^{nm}.
\end{equation}

In principle, the $\sigma$ fields should couple to the bundle describing the
`minimal' action of the gauge group.  If $S$ denotes the universal subbundle
on $G(k,n)$, and $\pi$ the projection from the gerbe to $G(k,n)$,
then we take the the $\sigma$ fields to couple to $\tilde{S}$
defined by
\begin{eqnarray}
\tilde{S} & = & \pi^* S \otimes \left( \pi^* \det S \right)^{-1/k}
\otimes \left( \pi^* \det S \right)^{1/(k (km+1))},
\\
& = & \pi^* S \otimes \left( \pi^* \det S \right)^{- m / (km+1)}.
\end{eqnarray}
(Note that since $\tilde{S}$ is of rank $k$,
the $(k (km+1))$th root is well-defined on a
${\mathbb Z}_{km+1}$ gerbe.)
We justify this identification by the fact that
\begin{equation}
\det \tilde{S} \: = \: \left( \det \pi^* S \right)^{1 / (km+1) }.
\end{equation}
This tells us that the $\sigma$ fields are coupling to a generator,
roughly speaking.

Now define $\tau_a = \sigma_a + m\sum_b \sigma_b$.  
In terms of $\tau_a$, equation~(\ref{eq:wgr:qh}) becomes
\begin{equation}   \label{eq:wgerbe:gkn}
    (-1)^{k-1} q = \tau_a^n \left(\det\tau\right)^{nm}.
\end{equation}

To interpret this result, we note that~(\ref{eq:wgerbe:gkn}) implies
\begin{equation}
    \left( \prod_a \tau_a^n \right) (\det \tau)^{knm} \: = \: (-)^{k(k-1)} q^k \: = \: q^k, 
\end{equation}
or more simply
\begin{equation}
    (\det \tau)^{n(km+1)} \: = \:  q^k, 
\end{equation}
hence
\begin{equation}
    (\det \tau)^{nm} \: = \: \xi^{m} \left[ q^k \right]^{m / (km+1)},
\end{equation}
for $\xi$ a $(km+1)$th root of unity.  Then, we can rewrite~(\ref{eq:wgerbe:gkn}) as
\begin{equation}   \label{eq:wgerbe:gkn:2}
    \tau_a^n \: = \: \xi^{-m} (-)^{k-1} q^{+1/(km+1)}.
\end{equation}

To interpret equation~(\ref{eq:wgerbe:gkn:2}),
recall that for ordinary Grassmannian, the nonequivariant relations in the quantum cohomology ring are
\begin{equation}
    (-1)^{k-1} q = \sigma_a^n,
\end{equation}
which give rise to the relation
\begin{equation}
    c(S) \, c(Q) = 1 + (-1)^{n-k} q.
\end{equation}
Here, the $\tau_a$ correspond to the Chern roots of
\begin{align}
    \tilde{S} \otimes (\det \tilde{S})^m  = \pi^* S.
\end{align}
Thus, 
 we see that for the ${\mathbb Z}_{km+1}$ gerbe on the Grassmannian, the relation~(\ref{eq:wgerbe:gkn:2}) should be interpreted as
\begin{equation}
    c(\pi^* S) \, c(\pi^* Q) = 1 + (-1)^{n-k} \xi^{-m}  q^{+1/(km+1)},
\end{equation}
or more simply, $km+1$ copies of the ordinary quantum cohomology ring of $G(k,n)$, with $\theta$ angle shifts
(encoded in the roots of unity $\xi$).  This is as expected from decomposition, and also correctly reduces to results for gerbes on projective spaces in the special case $k=1$.

\subsubsection{Quantum K theory}

Applying the same methods discussed earlier,
the Coulomb branch equation is given by
\begin{equation}  \label{eq:wgerbe:qk:1}
    (-1)^{k-1} q X_a^k = \left(1-X_a (\det X)^m\right)^n \prod_{b=1}^k \left[X_b \left(1-X_b (\det X)^m\right)^{nm}\right].
\end{equation}
We can quickly check that this expression has correct specializations:
\begin{itemize}
    \item When $k=1$, this specializes to the relation for ${\mathbb P}^{n-1}_{[\ell,\cdots,\ell]}$,
    a ${\mathbb Z}_{\ell}$ gerbe on the projective space ${\mathbb P}^{n-1}$,
        \begin{equation}
        (1-X^\ell)^{n \ell} = q, \quad {\rm with~} \ell = m+1,
        \end{equation}
        matching equation~(\ref{eq:qk:pn-1}) earlier.
        
    \item When $m=0$, this specializes to the relation~(\ref{eq:basicgkn:qk}) for ordinary Grassmannians.
\end{itemize}

Now, let us work out how to describe this in terms of decomposition.
Define
\begin{equation}
    M_a \: = \: X_a (\det X)^m, %\: \: \:
    %\beta \: = \: \det(M(1-M)^{nm}),
\end{equation}
%then taking a product over values of $a$, theCoulomb branch equation~(\ref{eq:wgerbe:qk:1}) implies
%\begin{equation}
%    q^k \: = \: \prod_{a=1}^k (1-M_a)^{n(mk+1)},
%\end{equation}
then 
%To simplify this, we note that 
the Coulomb branch equation~(\ref{eq:wgerbe:qk:1}) can be written 
\begin{equation} \label{eq:wgkn:1}
    (-1)^{k-1}q \prod_{b=1}^k \left(\frac{M_a (1-M_a)^{nm}}{M_b (1-M_b)^{nm}}\right) = (1-M_a)^{n(mk+1)} .
\end{equation}
%or more simply
%\begin{equation}
%    (-1)^{k-1} q M_a^k = (1-M_a)^n \beta .
%\end{equation}
Taking a product over values of $a$, this implies
%From~(\ref{eq:wgkn:1}), we have
\begin{equation}
    q^k \: = \: \prod_{a=1}^k (1-M_a)^{n(mk+1)},
\end{equation}
hence,
\begin{equation}
    \prod_{a=1}^k(1-M_a)^{nm} \: = \: q^{mk/(mk+1) } \zeta^m,
\end{equation}
where $\zeta$ is a $(mk+1)$th root of unity. Then we can write~(\ref{eq:wgerbe:qk:1}) as
\begin{equation} \label{eq:wgerbe:final}
    (-1)^{k-1} q^{1/(mk+1)} M_a^k \: = \: (1-M_a)^n  \zeta^m \det M,
\end{equation}
using the fact that, for example, $\det M = (\det X)^{1+km}$.

Comparing to the quantum K theory ring relations~(\ref{eq:basicgkn:qk}) for the ordinary Grassmannian $G(k,n)$,
we see that, 
as expected, the relation~(\ref{eq:wgerbe:final}) describes $km+1$ copies of the quantum K-theory ring relation of $G(k,n)$, indexed by the value of $\zeta$, in terms of the $M_a$, and as shifting the $X_a$ by $(km+1)$th roots of unity preserves the $M_a$, we see another $(km+1)$-fold ambiguity,
for altogether a decomposition into $(km+1)^2$ universes.

Finally, as a consistency check, let us take the $R \rightarrow 0$ limit and compare
to quantum cohomology.
We start from the Coulomb branch equations, repeated here
\begin{equation}
    (-1)^{k-1} q_{\rm 3d} X_a^k = \left(1-X_a (\det X)^m\right)^n \prod_{b=1}^k \left[X_b \left(1-X_b (\det X)^m\right)^{nm}\right].
\end{equation}
To get quantum cohomology, we take a small $R$ limit.  Expanding, we have
\begin{align}
    X_a &= \exp(-2\pi R \sigma_a) = 1- 2\pi R \sigma_a + \dots, \\
    \det X &=\exp\left(-2\pi R \sum_a \sigma_a\right) = 1- 2\pi R \sum_a \sigma_a + \dots,\\
    q_{\rm 3d} &= (2\pi R)^{n(mk+1)} q_{\rm 2d}.
\end{align}
Plugging these into the Coulomb branch equations and sending $R \to 0$, we obtain
equation~(\ref{eq:wgr:qh}), as expected.

\subsection{More general weighted Grassmannians}
\label{sect:genl-wgrass}

Next, for completeness, we consider a more general analogue of weighted projective spaces for Grassmannians,
and their quantum K theory.  Physically, these are described by 
a $U(k)$ gauge theory with $n$ chiral superfields, where the $i$th is in the $U(k)$ representation with highest weight
\begin{equation}
    [m_i+1, m_i, \dots, m_i].
\end{equation}
In special cases, this will be a gerbe.

The twisted one-loop effective superpotential (for the three-dimensional theory) is given by
\begin{align}
    W ={}& \frac{k}{2} \sum_a \left(\ln X_a\right)^2 - \frac{1}{2}\left(\sum_a \ln X_a\right)^2 + \ln (-1)^{k-1} q_{\mathrm{3d}} \sum_{a} \ln X_a \cr
    &+\sum_i\sum_a{\rm Li}_2 \left(X_a (\det X)^{m_i}\right).
\end{align}
The Coulomb branch equation is given by
\begin{equation}
    (-1)^{k-1} q_{\mathrm{3d}} X_a^k = \left(\det X\right) \prod_i \left[(1-X_a (\det X)^{m_i}) \prod_{b} (1-X_b (\det X)^{m_i})^{m_i}\right].
\end{equation}
Then taking the 2d limit, we obtain
\begin{equation}\label{eqn:qh-general-wts}
    (-1)^{k-1} q_{\rm 2d} =  \prod_i\left[ (\sigma_a + m_i \sum_c \sigma_c) \prod_b  (\sigma_b + m_i \sum_c \sigma_c)^{m_i}\right],
\end{equation}
where
\begin{equation}
    q_{\rm 3d}= (2\pi R)^{n + k(m_1+\dots+m_n)} q_\mathrm{2d}.
\end{equation}

\bigskip
Now, let $S$ denote a vector bundle associated to the ordinary fundamental of $U(k)$, of highest weight
$[1,0,\cdots,0]$, then the $\sigma_a$ couple to $S$, and $\sigma_a + m_i \sum_b \sigma_b$ couples to $S \otimes (\det S)^{m_i}$.

For the ordinary case, we symmetrized so that the expression is symmetric in $\sigma_a$'s. For the gerby case where all the $m_i$'s are equal to $m$, we defined $\tau_a = \sigma_a + m \sum_b \sigma_ b$ and made symmetrization in terms of $\tau_a$'s. Here, we define
\begin{equation}
    \tau_a^i = \sigma_a + m_i \sum_b \sigma_b,
\end{equation}
and we symmetrize over $\tau_a^i$ for each fixed $i = 1, \dots, n$. The Coulomb branch equations become
\begin{equation}
    (-1)^{k-1} q_{\rm 2d} = \tau_a^1 \tau_a^2 \dots \tau_a^n \prod_b (\tau_b^1)^{m_1} (\tau_b^2)^{m_2} \dots (\tau_b^n)^{m_n}.
\end{equation}

\subsection{Gerbes on flag manifolds}
\label{sect:gerbe:flag}

In this section, we will outline
\begin{equation}
    \prod_{i=1}^s {\mathbb Z}_{m_i k_i + 1}
\end{equation}
gerbes on a flag manifold $Fl(k_1, \cdots, k_s, n)$.

The GLSM for an ordinary flag manifold $Fl(k_1, k_2, \dots, k_s, n)$ is a $U(k_1) \times U(k_2) \times \cdots \times U(k_s)$ gauge theory with matter fields which are bifundamentals in the $(\mathbf{k_i}, \overline{\mathbf{k_{i+1}}})$ representation of $U(k_i) \times U(k_{i+1})$ for $i = 1, 2, \dots, s-1$, and $n$ fundamentals of $U(k_s)$ \cite{Donagi:2007hi}.

The GLSM for the desired gerbe on the 
flag manifold $Fl(k_1, k_2, \dots, k_s, n)$ is constructed as an generalization of the weighted Grassmannian.  Specifically, it is a $U(k_1) \times U(k_2) \times \cdots \times U(k_s)$ gauge theory, with chiral fields $\Phi^{(i)}$ transforming in the $U(k_i)$ representation with highest weight $[m_i+1, m_i, \dots, m_i]$ and in the $U(k_{i+1})$ representation with highest weight $[-m_{i+1}-1, -m_{i+1}, \dots, -m_{i+1}]$, for $i = 1, 2, \dots, s-1$.  There are also $n$ chiral fields $\Phi^{(s)}$ transforming in the $U(k_s)$ representation with highest weight $[m_s+1, m_s, \dots, m_s]$.

Then the pertinent twisted superpotential for the $i$th ($i = 1, 2, 3, \dots, s$) step is
\begin{align}
    W_i \: = \:{} & \frac{k_i}{2} \sum_{a=1}^{k_i} \left(\ln X_a^{(i)}\right)^2 - \frac{1}{2} \left(\sum_{a=1}^{k_i} \ln X_a^{(i)}\right)^2 + \left(\ln (-1)^{k_i - 1}q_i\right) \sum_{a=1}^{k_i} \ln X^{(i)}_a \cr
    &+ \sum_{a=1}^{k_i} \sum_{b=1}^{k_{i-1}} \mathrm{Li}_2\left(\frac{X_b^{(i-1)} (\det X^{(i-1)})^{m_{i-1}}}{X_a^{(i)} (\det X^{(i)})^{m_i}}\right) + \sum_{a=1}^{k_i} \sum_{b=1}^{k_{i+1}} \mathrm{Li}_2\left(\frac{X_a^{(i)} (\det X^{(i)})^{m_i}}{X_b^{(i+1)} (\det X^{(i+1)})^{m_{i+1}} }\right), \nonumber
\end{align}
with $k_0 = 0$ understood and $X_a^{(s+1)}$ being the equivariant parameters.

Let $Y_a^{(i)} = X^{(i)}_a (\det X^{(i)})^{m_i}$, the Coulomb branch equation is
\begin{align}
    (-1)^{k_i -1} q_i \left(Y_a^{(i)}\right)^{k_i} \prod_{b=1}^{k_{i-1}}\left[\left(1-\frac{Y^{(i-1)}_b }{Y^{(i)}_a }\right) \prod_{c=1}^{k_i}\left(1-\frac{Y_b^{(i-1)}}{Y_c^{(i)}}\right)^{m_i}\right]\cr = \left(\det Y^{(i)}\right) \prod_{b=1}^{k_{i+1}} \left[\left(1- \frac{Y_a^{(i)}}{Y_b^{(i+1)}}\right) \prod_{c=1}^{k_i} \left(1- \frac{Y_c^{(i)}}{Y_b^{(i+1)}}\right)^{m_i}\right].
\end{align}
For this, we obtain
\begin{equation}
    q_i^{k_i} \prod_{a=1}^{k_i} \prod_{b=1}^{k_{i-1}} \left(1 - \frac{Y_b^{(i-1)}}{Y_a^{(i)}}\right)^{m_ik_i + 1} \: =\: \prod_{a=1}^{k_i} \prod_{b=1}^{k_{i+1}} \left(1 - \frac{Y_a^{(i)}}{Y_b^{(i+1)}}\right)^{m_ik_i + 1}.
\end{equation}
Therefore, we obtain
\begin{equation}
    q_i^{\frac{k_i}{m_i k_i + 1}} \zeta_i \prod_{a,b}\left(1-\frac{Y_b^{(i-1)}}{Y_a^{(i)}}\right) \: = \: \prod_{a,b} \left(1 - \frac{Y_a^{(i)}}{Y_b^{(i+1)}}\right),
\end{equation}
where $\zeta_i$ is $(m_ik_i + 1)$th root of unity. Then, we can rewrite the Coulomb branch equation as
\begin{equation}
    (-1)^{k_i - 1} q_i^{\frac{1}{m_i k_i + 1}} (Y_a^{(i)})^{k_i} \prod_{b=1}^{k_{i-1}} \left(1 - \frac{Y_b^{(i-1)}}{Y_a^{(i)}}\right) = \left(\det Y^{(i)}\right) \prod_{b=1}^{k_{i+1}} \left(1 - \frac{Y_a^{(i)}}{Y_b^{(i+1)}}\right) \zeta_i^{m_i}.
\end{equation}
We see that, these relations describe $\prod_i (m_i k_i + 1)$ copies of the quantum K-theory ring relations of $Fl(k_1, k_2, \dots, k_s, n)$, indexed by $(\zeta_1, \zeta_2, \dots, \zeta_s)$, as generated by the $Y^{(i)}$'s, which are invariant under multiplication of the $X^{(i)}$ by $(k_im_i+1)$th roots of unity, for altogether a decomposition into
$\prod_i (m_i k_i + 1)^2$ universes.

We can take the $R \to 0$ limit, and obtain the quantum cohomology ring relations. We have
\begin{align}
    X_a^{(i)} &= \exp(-2\pi R \sigma_a^{(i)}) = 1 - 2\pi R \sigma_a^{(i)} + \dots\\
    \det X^{(i)} &=\exp\left(-2\pi R \sum_a \sigma_a^{(i)}\right) = 1- 2\pi R \sum_a \sigma_a^{(i)} + \dots\\
    q_i^{\rm 3d} &= (2\pi R)^{(k_{i+1} - k_{i-1})(m_ik_i+1)} q_i^{\rm 2d}.
\end{align}
Let
\begin{equation}
    \tau_a^{(i)} = \sigma_a^{(i)} + m_i \sum_b \sigma_b^{(i)},
\end{equation}
the quantum cohomology ring relations can be written as
\begin{align}
    (-1)^{k_i - 1} q_i^{\mathrm{2d}} \prod_{b=1}^{k_{i-1}}\left[\left(\tau_b^{(i-1)} - \tau_a^{(i)}\right) \prod_{c=1}^{k_i} \left(\tau_b^{(i-1)} - \tau_c^{(i)}\right)^{m_i} \right] \cr
    = \prod_{b=1}^{k_{i+1}} \left[\left(\tau_a^{(i)} - \tau_b^{(i+1)}\right) \prod_{c=1}^{k_i} \left(\tau_c^{(i)} - \tau_b^{(i+1)}\right)^{m_i}\right].
\end{align}
Finally, we have
\begin{equation}
    (-1)^{k_i -1} q_i^{1/(m_ik_i +1)} \prod_{b=1}^{k_{i-1}} \left(\sigma_b^{(i-1)} - \sigma_a^{(i)}\right) = \prod_{b=1}^{k_{i+1}} \left(\tau_a^{(i)} - \tau_b^{(i+1)}\right) \zeta_i^{m_i},
\end{equation}
for $\zeta_i$ a $(m_ik_i + 1)$th root of unity. Again, these describe $\prod_i (m_ik_i + 1)$ copies of the quantum cohomology ring relation of $Fl(k_1, k_2, \dots, k_s, n)$, indexed by the value of $\zeta_i$'s, as expected from decomposition for two-dimensional theories.

\subsection{More general levels: projective spaces}   \label{sect:glsm:badlevel}

So far, we have discussed three-dimensional GLSMs for gerbes with Chern-Simons terms chosen so as to get OPE rings matching quantum K theory in mathematics.
In this section, we will briefly outline projective spaces with more general levels, to outline some of the complications that can ensue.

Consider a GLSM for a gerby projective space ${\mathbb P}^n$, meaning a $U(1)$ gauge theory with $n+1$ chiral superfields of charge $\ell$,
and with Chern-Simons terms at level $k$.  Following \cite[equ'n (2.1)]{Gu:2020zpg},
the superpotential describing this theory is
\begin{equation}
    W \: = \: \frac{1}{2} \left( k + \ell^2 \frac{n+1}{2} \right) \left( \ln X \right)^2 
    \: + \: \left( \ln q \right) \left( \ln X \right) \: + \:
    \sum_{i=1}^{n+1} {\rm Li}_2\left( X^{\ell} \right).
\end{equation}
The equations of motion are
\begin{equation}   \label{eq:generallevel:proj1}
    \left( 1 - X^{\ell} \right)^{ \ell(n+1) } \: = \: q X^{K + \ell^2 (n+1)/2}.
\end{equation}

If we wanted to recover quantum K theory specifically, we would determined the Chern-Simons level from $U(1)_{-1/2}$ quantization, which would stipulate
\begin{equation}
    k \: = \: - \frac{1}{2} \sum_i (Q_i)^2 \: = \: - \ell^2 \frac{n+1}{2},
\end{equation}
where the $Q_i$'s are the gauge charges of the chiral superfields.  It is easy to
see that for this level, the equations of motion reduce to
\begin{equation}
    \left( 1 - X^{\ell} \right)^{ \ell(n+1) } \: = \: q,
\end{equation}
which we have discussed previously.

Now, suppose $k$ is more general.  Let us consider some cases.
\begin{itemize}
    \item First, suppose that $k$ is divisible by $\ell$:  $k = p \ell$ for some integer $p$.  In this case, from the general discussion of
    section~\ref{sect:predict}, there should be a $B {\mathbb Z}_{\ell}$ symmetry and a decomposition in the two-dimensional theory, and corresponding to that, 
    we can take an $\ell$th root of the equations of motion~(\ref{eq:generallevel:proj1}) to get
    \begin{equation}
        \left( 1 - X^{\ell} \right)^{n+1} \: = \: \xi q^{1/\ell} X^{p + \ell (n+1)/2},
    \end{equation}
    where $\xi$ is an $\ell$th root of unity.  For each choice of $\xi$, we get a different theory, and the equation above describes the classical solutions of that theory.
    \item If in addition, $k$ is divisible by $\ell^2$, then the equations of motion above are a polynomial in $X^{\ell}$.  If we write $p = \ell r$, and define $Y = X^{\ell}$, then the equations of motion become
    \begin{equation}
        \left( 1 - Y \right)^{n+1} \: = \: \xi q^{1/\ell} Y^{r + (n+1)/2},
    \end{equation}
    which are the vacua corresponding to the GLSM for an ordinary projective
    space ${\mathbb P}^n$ with level $r$.  Taking roots of $Y = X^{\ell}$ results in $\ell$ copies.  In other words, if $k = \ell^2 r$, then the equations of motion are the same as $\ell^2$ copies of those for the GLSM for ${\mathbb P}^n$ with Chern-Simons term at level $r$.  In short, a decomposition squared, as expected.

    However, it is essential for this last point that $k$ by divisible by $\ell^2$.
    If $k$ is only divisible by $\ell$, not $\ell^2$, then we do not get two orders of decomposition.
    \item For completeness, if $k$ is not divisible by $\ell$, but the gcd$(k, \ell) > 1$, then we can repeat a similar argument, in which we get at least an 
    order gcd$(k,\ell)$ decomposition, and potentially more if the Chern-Simons level has further divisibility properties.
\end{itemize}

We leave a thorough classification of all possibilities for future work.

\section{Conclusions}

In this paper we have discussed how decomposition \cite{Hellerman:2006zs} plays a role in three-dimensional gauge theories with one-form symmetries.  Although the three-dimensional theory itself does not decompose,
effective two-dimensional theories of parallel one-dimensional objects, or for that matter dimensional reductions, do decompose, in two separate ways.  As a result, if one starts with a theory with a
$B {\mathbb Z}_k$ one-form symmetry, the effective two-dimensional theory will decompose into, locally,
$k^2$ universes.  This was initially proposed in \cite{Gu:2021yek,Gu:2021beo}, and we have extended their analysis to more general cases (resulting in more complex decompositions).  This structure also immediately makes a prediction for quantum K theory rings, which are realized as OPE rings of parallel Wilson lines in three-dimensional theories.

In principle, the same ideas should apply in higher dimensions.
For example, parallel surfaces in four-dimensional gauge theories with one-form symmetries should also exhibit
decomposition in their OPEs, in multiple ways, even though the theory as a whole does not decompose, as outlined in the introduction.
We leave this for future work.  similar ideas should also apply in theories with gauged trivially-acting noninvertible symmetries, as discussed in e.g.~\cite{Perez-Lona:2023djo}.

\section{Acknowledgements}

We would like to thank W.~Gu, A.~Perez-Lona, I.~Melnikov, L.~Mihalcea, T.~Pantev, 
H.-H.~Tseng, W.~Xu, and X.~Yu for useful comments.  We would also like to thank the referee for suggesting a generalization of the analysis in the original draft.
E.S. was partially supported by NSF grant PHY-2310588.

\appendix

\section{Bundles on stacks and gerbes}  \label{app:bundles-gerbes}

Deligne-Mumford stacks can typically be presented as $[X/G]$, where $G$ is any group (not necessarily finite), with any
action on $X$ (not necessarily effective).  The stack $[X/G]$ is said to be a $K$-gerbe if a subgroup $K \subset G$
acts trivially on $X$.

The cohomology of the stack ${\mathfrak X} = [X/G]$ is most naturally defined on the inertia stack $I_{\mathfrak X}$.
Intuitively, the inertia stack is the zero-momentum part of the loop space of ${\mathfrak X}$, and as such,
has one component which is a copy of ${\mathfrak X}$, plus other components (due to the existence of automorphisms encoded
in ${\mathfrak X}$.  Each component is a copy of a substack of ${\mathfrak X}$.
Those components are labelled by automorphisms $\alpha$.  The group generated by any
automorphism $\alpha$, call it $\langle \alpha \rangle$, is cyclic.

For one example, suppose ${\mathfrak X} = [{\mathbb C}^2/{\mathbb Z}_2]$, with the ${\mathbb Z}_2$ acting by sign flips.
This has one fixed point, at the origin of the plane ${\mathbb C}^2$.  In this case,
\begin{equation}
    I_{\mathfrak X} \: = \: [{\mathbb C}^2/{\mathbb Z}_2] \, \coprod \, [{\rm point}/{\mathbb Z}_2].
\end{equation}
The second component is associated with an order-two automorphism.

For another example, suppose ${\mathfrak X} = [X/{\mathbb Z}_k]$ where all of ${\mathbb Z}_k$ acts trivially.
In that case,
\begin{equation}
    I_{\mathfrak X} \: = \: \coprod_{m=0}^{k-1} [X/{\mathbb Z}_k].
\end{equation}

Let $\pi: I_{\mathfrak X} \rightarrow {\mathfrak X}$ denote the projector whose restriction to any component is the
projection onto that component.  We denote the restriction of $\pi$ to the component $\lambda$ by 
$\pi_{\lambda}$.

Let $E \rightarrow {\mathfrak X}$ be a vector bundle.
A sheaf or bundle on the stack ${\mathfrak X} = [X/G]$, 
is precisely the same as a $G$-equivariant sheaf or bundle on $X$, the covering space,
so $E$ is the same as a $G$-equivariant bundle on $X$.

On each component of $I_{\mathfrak X}$, $\pi_{\lambda}^* E$ will decompose into eigenbundles of the action of the
stabilizer $\alpha(\lambda)$:
\begin{equation}
    \pi_{\lambda}^* E |_{\lambda} \: = \: \bigoplus_{\chi} E_{\lambda, \chi},
\end{equation}
where $\chi$ is a character of the stabilizer $\alpha(\lambda)$.
One defines ch$^{\rm rep}(E)$ over a component of $I_{\mathfrak X}$ to be
\begin{equation}
    {\rm ch}^{\rm rep}(E)|_{\lambda} \: = \: \bigoplus_{\chi} {\rm ch}( E_{\lambda, \chi}) \otimes \chi, 
\end{equation}
where ch denotes the ordinary Chern character in equivariant cohomology.  The reader should note that, curiously,
ch$^{\rm rep}$ is a complex-valued cohomology class.

\end{document}